\documentclass[prb,twocolumn,showpacs,aps,superscriptaddress,floatfix]{revtex4}
\usepackage{amsmath}
\usepackage{amssymb}
\usepackage{amsfonts}
\usepackage{bm}
\usepackage{graphicx}
\usepackage{color}
\usepackage[svgnames]{xcolor}
\usepackage{soul}
\usepackage{hyperref}
\usepackage{verbatim}
\graphicspath{{FIGS/}}

\begin{document}

\title{Coexistence of nematic superconductivity and spin density wave in magic-angle twisted bilayer graphene}

\author{A.O. Sboychakov}
\affiliation{Institute for Theoretical and Applied Electrodynamics, Russian
Academy of Sciences, Moscow, 125412 Russia}

\author{A.V. Rozhkov}
\affiliation{Institute for Theoretical and Applied Electrodynamics, Russian
Academy of Sciences, Moscow, 125412 Russia}

\author{A.L. Rakhmanov}
\affiliation{Institute for Theoretical and Applied Electrodynamics, Russian
Academy of Sciences, Moscow, 125412 Russia}

\begin{abstract}
We argue that doped twisted bilayer graphene with magical twist angle can
become superconducting. In our theoretical scenario, the superconductivity
coexists with the spin-density-wave-like ordering. Numerical mean-field
analysis demonstrates that the spin-density-wave order, which is much
stronger than the superconductivity, leaves parts of the Fermi surface
ungapped. This Fermi surface serves as a host for the superconductivity.
Since the magnetic texture at finite doping breaks the point group of the
twisted bilayer graphene, the stabilized superconducting order parameter is
nematic. We also explore the possibility of a purely Coulomb-based
mechanism of superconductivity in the studied system. The screened Coulomb
interaction is calculated within the random phase approximation. It is
shown that near the half-filling the renormalized Coulomb repulsion indeed
induces the superconducting state, with the order parameter possessing two
nodes on the Fermi surface. We estimate the superconducting transition
temperature, which turns out to be very low. The implications of our
proposal are discussed.
\end{abstract}


\date{\today}

\maketitle

\section{Introduction}

The discovery of Mott insulating
states~\cite{NatureMott2018,MottSCNature2019}
and
superconductivity~\cite{NatureSC2018,MottSCNature2019}
in magic angle twisted bilayer graphene (MAtBLG) has attracted great
attention to this material. In twisted bilayer graphene (tBLG) one graphene
layer is rotated with respect to another one by a twist angle $\theta$. The
twisting produces a moir{\'{e}} pattern and superstructure in the system.
The low-energy electronic structure of tBLG is substantially modified in
comparison to single-layer, AA-stacked, and AB-stacked bilayer
graphene~\cite{ourBLGreview2016}.
For small
$\theta \sim 1^{\circ}$,
the low-energy single-electron spectrum consists of eight (if spin degree of freedom is accounted for) flat bands
separated from lower and higher dispersive bands by energy
gaps~\cite{ourBLGreview2016,dSPRB,NonAbelianGaugePot}.
The width of the low-energy bands (which is about several meV) has a
minimum at
$\theta = \theta_c$,
where
$\theta_c$
is the so-called magic angle
$\theta_c \sim 1^{\circ}$.

The existence of the flat bands makes MAtBLG very susceptible to
interactions. The interactions lead to the appearance of Mott insulating
states when carrier doping per superlattice cell $n$ is an integer. The
authors of
Ref.~\onlinecite{NatureMott2018}
observed the insulating states in transport measurements near the
neutrality point (zero doping) and at doping corresponding to
$n=\pm2$
extra charges per supercell. In similar experiments in
Ref.~\onlinecite{MottSCNature2019}
the authors observed Mott states at doping corresponding to
$n=0$,
$n=\pm1$,
$n=\pm2$,
and
$n=\pm3$.
The nature of the insulating ground states is under
discussion~\cite{Philips2018,PhysRevB.98.081102,ChiralSDW_SC2018,
AFMMottSC2019, OurtBLGPRB2019, Nematic, OurJETPLetters2022,
OurPhaSepJETPL2020, cea2020band, IVCMonteCarloPRX2022, U4PRL2022,
IncommKekuleSpiralPRL2022, FMorderPRL2019}.
Several types of ordering, such as spin-density wave (SDW)
states~\cite{ChiralSDW_SC2018,AFMMottSC2019,OurtBLGPRB2019,Nematic,
OurJETPLetters2022},
ferromagnetic
state~\cite{FMorderPRL2019},
and other symmetry-broken
phases~\cite{cea2020band,IVCMonteCarloPRX2022,IncommKekuleSpiralPRL2022,
U4PRL2022}
have been proposed to be the ground state of the system.

Besides Mott insulating states, the authors of
Ref.~\onlinecite{NatureSC2018}
observed
on the doping-temperature ($n,T$)
plane
two superconductivity domes
located slightly below and slightly above half-filling,
$n=-2$.
In other experiments\cite{MottSCNature2019},
the superconductivity domes have been observed close to
$n=-2$,
$n=0$,
and
$n=\pm1$.

Theory of the superconductivity in the MAtBLG has been developed in many
papers, see, e.g.,
Refs.~\onlinecite{PhononSCMcDonald2018,PhononSCPRL2019,PhysRevB.97.235453,
ChiralSDW_SC2018,KLSC2019,AFMMottSC2019,PhysRevB.99.121407,
TheorySCNature2022,PhononSCPRB2022,PhysRevB.107.024509}.
Different mechanisms, including
phonon~\cite{PhononSCMcDonald2018,PhononSCPRL2019,PhononSCPRB2022}
and
electronic~\cite{PhysRevB.97.235453,ChiralSDW_SC2018,KLSC2019,AFMMottSC2019,
PhysRevB.99.121407},
are under discussion. The symmetry of the superconducting order parameter
is debated as well. All cited works suggest that the superconductivity does
not coexist with any non-superconducting order parameter (with the
exception of
Ref.~\onlinecite{PhysRevB.99.121407},
where such a possibility is considered).

In our previous
papers~\cite{OurtBLGPRB2019,Nematic,OurPhaSepJETPL2020,OurJETPLetters2022}
we studied the non-superconducting order in MAtBLG assuming that the SDW is
the ground state of the system. We showed that the SDW is stable in the
doping range
$-4 < n < 4$.
This allowed us to explain the behavior of the conductivity versus doping
(of course, that theory is applicable only outside of the regions where
superconductivity was observed). We showed also that at finite doping the
point symmetry of the SDW state is reduced, and electronic nematicity
emerges~\cite{Nematic}.
The latter is indeed confirmed by
experiment~\cite{MottNematicNature2019,KerelskyNematicNature2019}.

In the present paper we focus on the superconductivity. We consider the
doping range close to half-filling,
$n=-2$.
We assume here that the superconductivity coexists, but does not compete,
with the SDW phase. This expectation is based on the observation that the
SDW order, with its characteristic energy of several tens of meV, is much
stronger than the superconductivity, whose transition temperature is as low
as
$T_c=1.7$\,K.
Under such circumstances, theoretical justification for the coexistence
relies on the presence of a Fermi surface that remains in MAtBLG even when
SDW order is established.

Additionally, we investigate a non-phonon mechanism of superconductivity
for MAtBLG. Our proposal relies on the renormalized Coulomb potential,
which we calculate using the random phase approximation (RPA). It will be
demonstrated that the screened Coulomb interaction can indeed stabilize the
superconductivity coexisting with the SDW. The superconducting order
parameter has two nodes on the Fermi surface, similar to a $p$-wave order.
However, as the SDW spin texture breaks several symmetries, the common
order-parameter classification into $s$-wave, $p$-wave, etc., does not
apply. The estimated critical temperature turns out to be significantly
smaller than the experimentally observed values. This discrepancy is
discussed from the theory standpoint. Possible reasons behind it are
analyzed.

The paper is organized as follows. The geometry of the system under study
is briefly described in
Sec.~\ref{Geometry}.
In
Section~\ref{Model}
we formulate our model and describe the structure of the SDW order
parameter.
Section~\ref{PolarizationOperator}
is devoted to the static polarization operator and the renormalized Coulomb
potential. In
Section~\ref{superconductivity}
we derive the self-consistency equation for the superconducting order
parameter coexisting with the SDW order. We describe the property of the
superconducting order and obtain an estimate for
$T_c$.
Discussion and conclusions are presented in
Section~\ref{Conclusions}.

\section{Geometry of twisted bilayer graphene}
\label{Geometry}

In this Section we recap several basic facts about the geometry of the tBLG
that are important for further consideration (for more details, see, e.g.,
reviews
Refs.~\onlinecite{MeleReview,ourBLGreview2016}).
Each graphene layer in tBLG forms a hexagonal honeycomb lattice that can be
split into two triangular sublattices,
${\cal A}$
and
${\cal B}$.
The coordinates of atoms in layer~1 on sublattices
${\cal A}$
and
${\cal B}$
are
\begin{equation}
\mathbf{r}_{\mathbf{n}}^{1{\cal A} }
=
\mathbf{r}_{\mathbf{n}}^{1}\equiv n\mathbf{a}_1+m\mathbf{a}_2\,,\;\;
\mathbf{r}_{\mathbf{n}}^{1{\cal B} }
=
\mathbf{r}_{\mathbf{n}}^{1}+\bm{\delta}\,,
\end{equation}
where
$\mathbf{n}=(n,\,m)$
is an integer-valued vector,
$\mathbf{a}_{1,2}=a(\sqrt{3},\mp1)/2$
are the primitive vectors,
$\bm{\delta}=(\mathbf{a}_1+\mathbf{a}_2)/3=a(1/\sqrt{3},0)$
is a vector connecting two atoms in the same unit cell, and
$a=2.46$\,\AA\
is the lattice constant of graphene. Atoms in layer~2 are located at
\begin{equation}
\mathbf{r}_{\mathbf{n}}^{2{\cal B}}
=
\mathbf{r}_{\mathbf{n}}^{2}
\equiv
d\mathbf{e}_z+n\mathbf{a}_1'+m\mathbf{a}_2'\,,
\quad
\mathbf{r}_{\mathbf{n}}^{2{\cal A}}
=
\mathbf{r}_{\mathbf{n}}^{2}-\bm{\delta}'\,,
\end{equation}
where
$\mathbf{a}_{1,2}'$
and
$\bm{\delta}'$
are the vectors
$\mathbf{a}_{1,2}$
and
$\bm{\delta}$,
rotated by the twist angle
$\theta$.
The unit vector along the
$z$-axis is
$\mathbf{e}_z$,
the interlayer distance is
$d=3.35$\,\AA. The limiting case
$\theta=0$
corresponds to the AB stacking.

\begin{figure}[t]
\centering
\includegraphics[width=0.99\columnwidth]{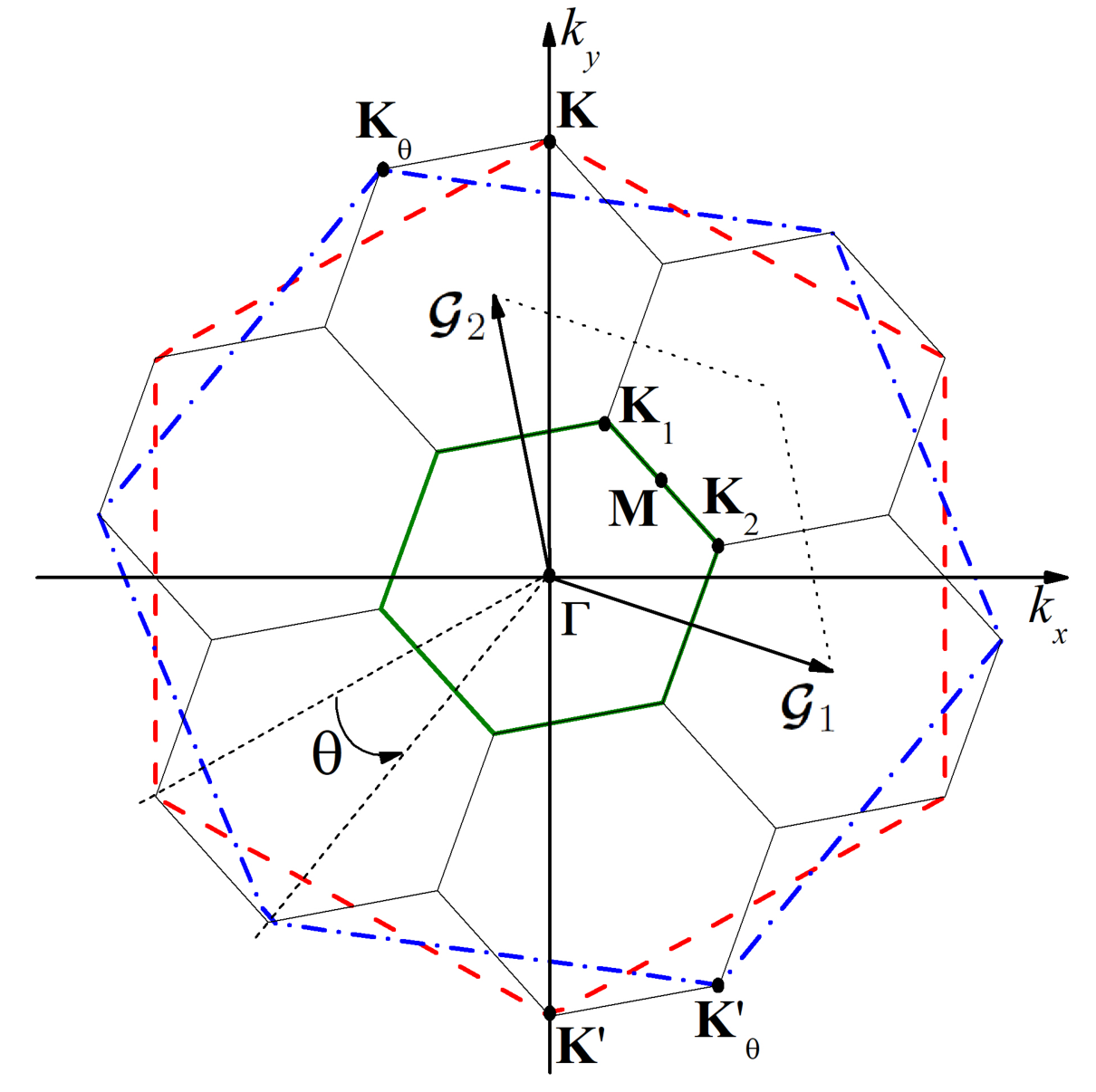}
\caption{Reciprocal space geometry of tBLG for
$\theta \approx 21.79^{\circ}$
($m_0=1$,
$r=1$).
The figure presents Brillouin zones of layers~1 and~2 (large red and blue
hexagons), as well as the Brillouin zone of the superlattice (small thick
green hexagon). Reciprocal vectors of the superlattice
($\bm{{\cal G}}_{1}$
and
$\bm{{\cal G}}_{2}$),
Dirac points of layer~1
($\mathbf{K}$
and
$\mathbf{K}'$)
and layer~2
($\mathbf{K}_{\theta}$
and
$\mathbf{K}'_{\theta}$),
as well as high symmetry points of the reduced Brillouin zone
($\bm{\Gamma}$,
$\mathbf{M}$,
$\mathbf{K}_{1,2}$)
are also shown.
\label{fig::FigTBLGSCBZ}
}
\end{figure}

Twisting produces moir{\'{e}}
patterns~\cite{ourBLGreview2016},
which can be seen as alternating dark and bright regions in STM images.
Measuring the moir{\'{e}} period $L$, one can extract the twist angle using
the formula
$L=a/[2\sin(\theta/2)]$.
Moir{\'{e}} patterns exist for arbitrary twist angles. If the twist angle
satisfies the relationship
\begin{equation}
\label{comtheta}
\cos\theta=\frac{3m_0^2+3m_0r+r^2/2}{3m_0^2+3m_0r+r^2}\,,
\end{equation}
where $m_0$ and $r$ are co-prime positive integers, it is called
commensurate. For commensurate $\theta$'s a superstructure emerges, and the
sample splits into a periodic lattice of finite supercells. The majority of
theoretical papers assume the twist angle to be the commensurate one, since
only in this case one can work with Bloch waves and introduce the
quasimomentum. For the commensurate structure described by $m_0$ and $r$,
the superlattice vectors are
\begin{equation}
\label{R12}
\mathbf{R}_1\!=\!m_0\mathbf{a}_1+(m_0+r)\mathbf{a}_2,\,
\mathbf{R}_2\!=\!-(m_0+r)\mathbf{a}_1+(2 m_0+r)\mathbf{a}_2,
\end{equation}
if
$r \neq 3n$
($n$ is an integer), or
\begin{equation}
\mathbf{R}_1=(m_0+n)\mathbf{a}_1+n\mathbf{a}_2,\,
\mathbf{R}_2=-n\mathbf{a}_1+(m_0+2n)\mathbf{a}_2,
\end{equation}
if
$r=3n$.
The number of graphene unit cells inside a supercell is
$N_{\rm sc}=(3m_0^2+3m_0r+r^2)/g$
per layer. The parameter $g$ in the latter expression is equal to unity
when
$r\neq3n$.
Otherwise, it is
$g=3$.

The superlattice cell of the structure with $m_0$ and $r$ contains
$r^2$
moir{\'{e}} cells if
$r\neq3n$,
or
$r^2/3$
moir{\'{e}} cells otherwise. When
$r=1$,
the superlattice cell coincides with the moir{\'{e}} cell. In the present
paper we consider only such structures. When $\theta$ is small enough, the
superlattice cell can be approximately described as consisting of regions
with almost AA, AB, and BA
stackings~\cite{dSPRB,ourBLGreview2016}.

The reciprocal lattice primitive vectors for layer~1 (layer~2) are denoted by
$\mathbf{b}_{1,2}$
($\mathbf{b}_{1,2}'$). For layer~1 one has
$\mathbf{b}_{1,2}=(2\pi/\sqrt{3},\mp 2\pi )/a$,
while
$\mathbf{b}_{1,2}'$
are connected to
$\mathbf{b}_{1,2}$
by a rotation of an angle
$\theta$.
Using the notation
$\bm{{\cal G}}_{1,2}$
for the primitive reciprocal vectors of the superlattice, the following identities in the reciprocal space are valid:
\begin{equation}
\mathbf{b}_1'=\mathbf{b}_1+r(\bm{{\cal G}}_{1}+\bm{{\cal G}}_{2})\,,\;
\mathbf{b}_2'=\mathbf{b}_2-r\bm{{\cal G}}_{1}\,,
\end{equation}
if
$r\neq3n$,
or
\begin{equation}
\mathbf{b}_1'=\mathbf{b}_1+n(\bm{{\cal G}}_{1}+2\bm{{\cal G}}_{2})\,,\;
\mathbf{b}_2'=\mathbf{b}_2-n(2\bm{{\cal G}}_{1}+\bm{{\cal G}}_{2})\,,
\end{equation}
if
$r=3n$.

Each graphene layer in tBLG has a hexagonal Brillouin zone. The Brillouin
zone of the layer~2
is rotated in momentum space with respect to the Brillouin zone of layer~1
by the twist angle $\theta$. The Brillouin zone of the superlattice
(reduced Brillouin zone, RBZ) is also hexagonal but smaller in size. It can
be obtained by
$N_{\rm sc}$-times
folding of the Brillouin zone of the layer~1 or~2. Two non-equivalent Dirac
points of layer~1 can be chosen as
$\mathbf{K}=(0,4\pi/(3a)), \mathbf{K}'=-\mathbf{K}$.
The Dirac points of layer~2
are
$\mathbf{K}_{\theta}=4\pi(-\sin\theta,\cos\theta)/(3a)$,
$\mathbf{K}'_{\theta}=-\mathbf{K}_{\theta}$.
Band folding translates these four Dirac points to the two Dirac points of the superlattice,
$\mathbf{K}_{1,2}$.
Thus one can say that the Dirac points of the superlattice are doubly
degenerate. Points
$\mathbf{K}_1$
and
$\mathbf{K}_2$
can be expressed via vectors
$\bm{{\cal G}}_{1,2}$
as
\begin{equation}
\mathbf{K}_1=\frac{1}{3}(\bm{{\cal G}}_{1}+2\bm{{\cal G}}_{2})\,,\quad
\mathbf{K}_2=\frac{1}{3}(2\bm{{\cal G}}_{1}+\bm{{\cal G}}_{2})\,.
\end{equation}
A typical picture illustrating these three Brillouin zones, the vectors
$\bm{{\cal G}}_{1,2}$,
as well as main symmetrical points is shown in
Fig.~\ref{fig::FigTBLGSCBZ}.

\section{Model Hamiltonian}
\label{Model}

We start with the following Hamiltonian of the tBLG:
\begin{eqnarray}
\label{H}
H
&=&\!\!\!
\sum_{{\mathbf{nm}ij\atop ss'\sigma}}\!
	t(\mathbf{r}_{\mathbf{n}}^{is};\mathbf{r}_{\mathbf{m}}^{js'})
	d^{\dag}_{\mathbf{n}is\sigma}
	d^{\phantom{\dag}}_{\mathbf{m}js'\sigma}
+
U\!\sum_{{\mathbf{n}is}}\!
	n_{\mathbf{n}is\uparrow}n_{\mathbf{n}is\downarrow}+
\nonumber
\\
&&\frac12\!\mathop{{\sum}'}_{{\mathbf{nm}ij\atop ss'\sigma\sigma'}}\!
	V(\mathbf{r}_{\mathbf{n}}^{is}-\mathbf{r}_{\mathbf{m}}^{js'})
	n_{\mathbf{n}is\sigma}n_{\mathbf{m}js'\sigma'}\,.
\end{eqnarray}
In this expression
$d^{\dag}_{\mathbf{n}is\sigma}$
($d^{\phantom{\dag}}_{\mathbf{n}is\sigma}$)
are the creation (annihilation) operators of the electron with spin
$\sigma$\,($=\uparrow,\downarrow$)
at the unit cell $\mathbf{n}$ in the layer
$i$\,($=1,2$)
in the sublattice
$s$\,($={\cal A,B}$),
while
$n_{\mathbf{n}is\sigma}
=
d^{\dag}_{\mathbf{n}is\sigma}d^{\phantom{\dag}}_{\mathbf{n}is\sigma}$.
The first term in
Eq.~\eqref{H}
is the single-particle tight-binding Hamiltonian with
$t(\mathbf{r}_{\mathbf{n}}^{is};\mathbf{r}_{\mathbf{m}}^{js'})$
being the amplitude of the electron hopping from site in the position
$\mathbf{r}_{\mathbf{m}}^{js'}$
to the site in the position
$\mathbf{r}_{\mathbf{n}}^{is}$.
The second term in
Eq.~\eqref{H}
describes the on-site (Hubbard) interaction of electrons with opposite
spins, while the last term corresponds to the intersite Coulomb
interaction. [The prime near the last sum in
Eq.~\eqref{H}
means that elements with
$\mathbf{r}_{\mathbf{n}}^{is}=\mathbf{r}_{\mathbf{m}}^{js'}$
should be excluded.]

\begin{figure}[t]
\centering
\includegraphics[width=0.99\columnwidth]{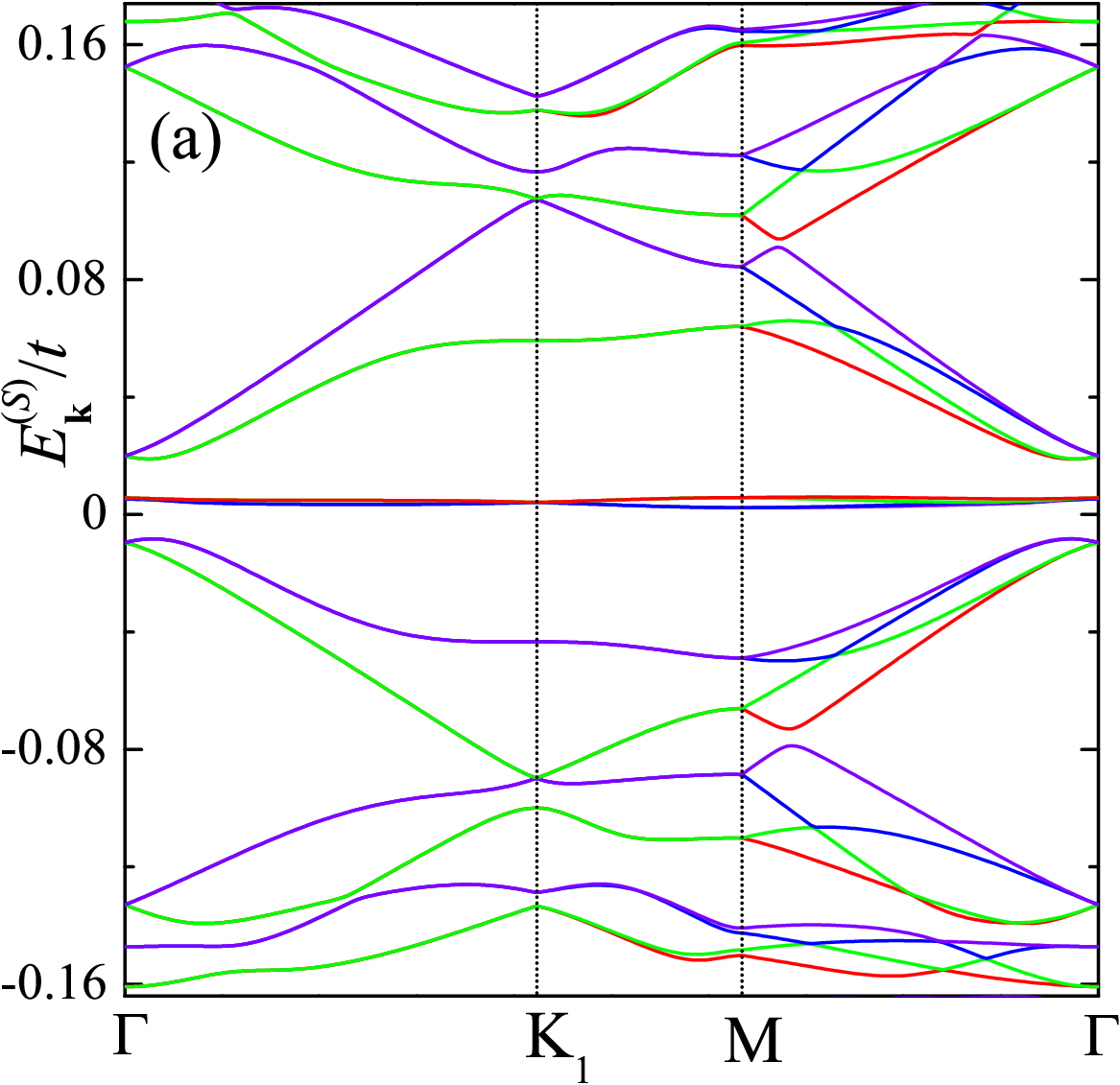}\\
\includegraphics[width=0.99\columnwidth]{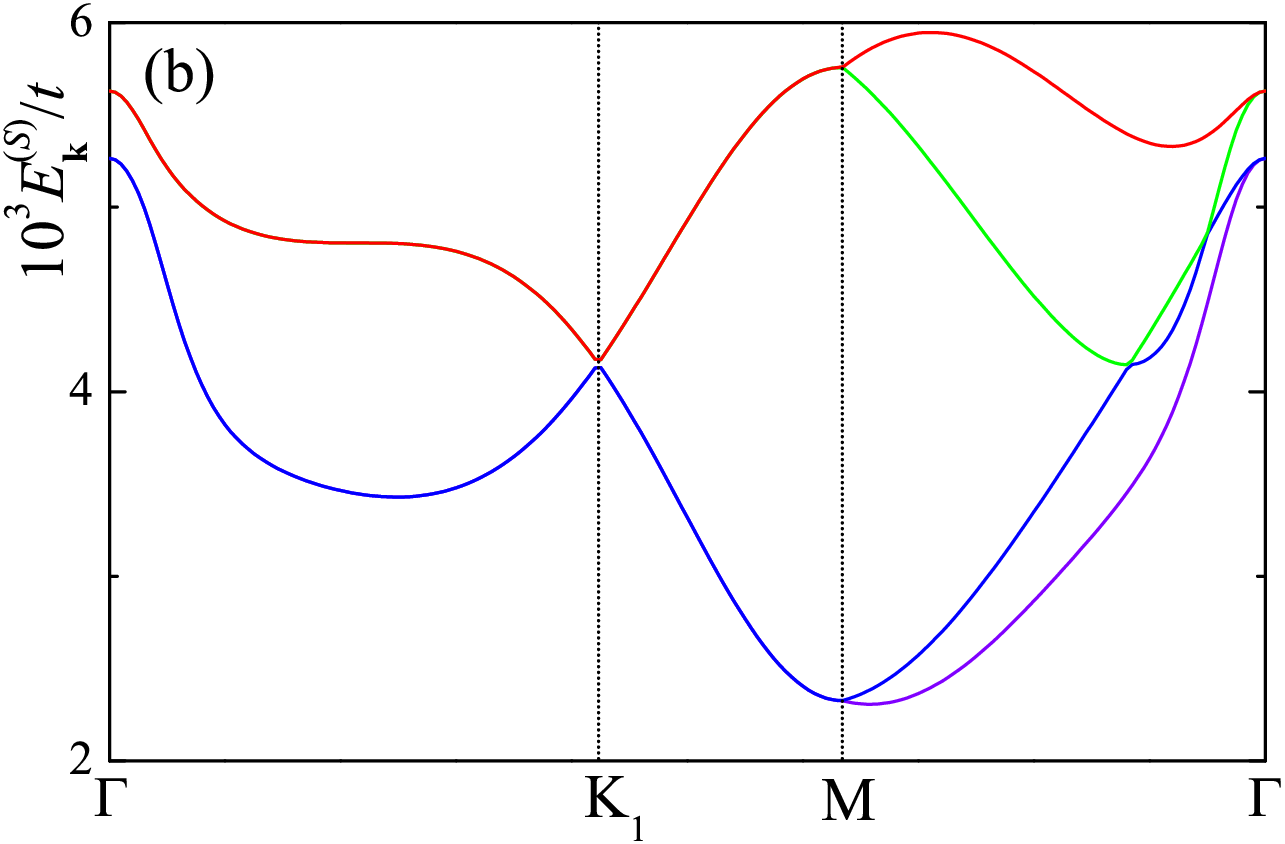}
\caption{(a) Energy spectrum of MAtBLG at vanishing interactions. The flat
bands appears as a bundle of almost nondispersive horizontal lines at
$E_{\bf k}^{(S)} \approx 4\times 10^{-3} t$.
The gaps separating the flat bands from the dispersive bands are clearly
visible. The data in panel~(b) show that the flat bands do have finite
dispersion as well. Their width $W$ can be approximately estimated as
$W \approx 9.4$\,meV
\label{FigSpec}
}
\end{figure}

\subsection{Single-particle spectrum of MAtBLG}

Let us consider first the single-particle properties of the MAtBLG. If we
neglect interactions, the electronic spectrum of the system is obtained by
diagonalization of the first term of the
Hamiltonian~\eqref{H}.
The result depends on the parametrization of the hopping amplitudes
$t(\mathbf{r}_{\mathbf{n}}^{is};\mathbf{r}_{\mathbf{m}}^{js'})$.
In this paper we keep only nearest-neighbor terms for the intralayer
hopping. The corresponding amplitude is
$t=-2.57$\,eV.

As for the interlayer hopping amplitudes, we explored several
parametrization schemes, all of which deliver qualitatively similar
results. The results presented below correspond to the parametrization II.B
of
Ref.~\onlinecite{Nematic}.
This parametrization, initially proposed in
Ref.~\onlinecite{Tang},
takes into account the environment dependence of the hopping. That is, the
electron hopping amplitude connecting two atoms at positions
$\mathbf{r}$
and
$\mathbf{r}'$
depends not only on the difference
$\mathbf{r}-\mathbf{r}'$,
but also on positions of other atoms in the lattice. Extra flexibility of
the formalism becomes useful when the tunneling between
$\mathbf{r}$
and
$\mathbf{r}'$
is depleted by nearby atoms, which act as obstacles to a tunneling
electron. For the tBLG, the parametrization~II.B was used in
Refs.~\onlinecite{Pankratov1,ourTBLG,ourTBLG2017},
among other papers. This parametrization can correctly reproduce the
Slonczewski-Weiss-McClure parametrization scheme in the limiting case of
the AB bilayer graphene
($\theta=0$).

Once a specific parametrization is chosen, the single-electron Hamiltonian
may be diagonalized and its energy spectrum may be found. For
parametrization chosen, the magic angle superstructure is
$(m_0,\,r)=(17,\,1)$,
which corresponds to the magic angle
$\theta_c=1.89^{\circ}$.

To execute the Hamiltonian diagonalization, one must introduce the
quasimomentum representation. To this end, we define new electronic
operators
$d^{\phantom{\dag}}_{\mathbf{pG}is\sigma}$
by the following relation
\begin{eqnarray}
\label{dpG}
d^{\phantom{\dag}}_{\mathbf{pG}is\sigma}=\frac{1}{\sqrt{{\cal{N}}}}\sum_{\mathbf{n}}
\exp{[-i(\mathbf{p}+\mathbf{G})\mathbf{r}_{\mathbf{n}}^{i}]}d_{\mathbf{n}is\sigma}\,,
\end{eqnarray}
where
${\cal N}$
is the number of graphene unit cells in the sample in one layer, the
momentum
$\mathbf{p}$
lies in the first Brillouin zone of the superlattice, while
$\mathbf{G}=n\bm{{\cal G}}_1+m\bm{{\cal G}}_2$
is the reciprocal vector of the superlattice confined to the first
Brillouin zone of the $i$th~layer. The number of
$\mathbf{G}$'s
satisfying the latter requirement is equal to
$N_{\rm sc}$
for each graphene layer.

In the quasimomentum representation, for a specific quasimomentum
${\bf p}$,
the single-electron Hamiltonian is a bilinear of the fermionic operators,
characterized by a
$4N_{\rm sc} \times 4N_{\rm sc}$
matrix (one such matrix per spin projection). Diagonalizing this matrix
numerically, one finds the single-electron spectrum of tBLG. The low-energy part of the spectrum is shown in
Fig.~\ref{FigSpec}.
In
this figure
we see four flat bands
$E_{0\mathbf{k}}^{({\cal S})}$
separated from lower and higher dispersive bands by the energy gaps of the
order of $30$\,meV. The width of the flat bands $W$ as a function of the
twist angle $\theta$ has a minimum
$W=9.4$\,meV
at the magic angle.

Unlike undoped graphene and undoped AB~bilayer, which both have Fermi
points, tBLG at low $\theta$ is a
metal~\cite{ourTBLG}
even at no doping. The four flat bands cross the Fermi level forming
multi-component Fermi surface, see Fig.~8 in
Ref.~\onlinecite{ourTBLG}.
The shape of the Fermi surface components depend on the specific model of
the interlayer hopping and on the doping level $n$.

\subsection{SDW order parameters}

The system having flat bands intersecting the Fermi level is very
susceptible to interactions. Interactions spontaneously break symmetries of
the single-particle Hamiltonian generating an order parameter. Neglecting
first a possibility of the superconducting state, we assume that this order
parameter is the SDW. This choice is not arbitrary. It was shown in many
papers (see, e.g.,
Refs.~\onlinecite{dSPRB,NonAbelianGaugePot,ourTBLG})
that at small twist angles, electrons on the Fermi level occupy mainly the
regions with almost perfect AA stacking within a supercell. At the same
time, it was demonstrated
theoretically~\cite{AAPRL, graph_phasep2013, aa_graph2013, akzyanov_aa2014}
that the ground state of AA stacked bilayer graphene is antiferromagnetic.
For this reason we believe that the ground state of MAtBLG possesses an
SDW-like order parameter.

The SDW order parameter is a multicomponent one. First, it contains on-site
terms of the form
\begin{eqnarray}
\label{Deltania}
\Delta_{\mathbf{n}is}
=
U\langle
	d^{\dag}_{\mathbf{n}is\uparrow}
	d^{\phantom{\dag}}_{\mathbf{n}is\downarrow}
\rangle\,,
\end{eqnarray}
with the on-site interaction $U$ serving as a proportionality coefficient.
For our calculations we assign
$U=2t$.
This value of $U$ is somewhat smaller than the critical
$U_c=2.23t$
above which single-layer graphene spontaneously enters a mean-field
antiferromagnetic
state~\cite{MF_Uc_sorella1992}.
Thus our Hubbard interaction is rather strong, but not strong enough to
open a gap in single-layer graphene.

Next, we include an intralayer nearest-neighbor SDW order parameter, which
is defined on links connecting nearest neighbor atoms in the same layer.
In a graphene layer, each atom in one sublattice has three nearest
neighbors belonging to the opposite sublattice: an atom on sublattice 
${\cal B}$
(sublattice
${\cal A}$)
has three nearest neighbors on sublattice
${\cal A}$
(sublattice
${\cal B}$).
For this reason we consider three types of intralayer nearest-neighbor
order parameters,
$A^{(\ell)}_{\mathbf{n}i\sigma}$
($\ell=1,\,2,\,3$),
corresponding to three different links connecting the nearest-neighbor
sites. These order parameters are defined as follows
\begin{equation}
\label{Anis}
A^{(\ell)}_{\mathbf{n}i\sigma}=V_{\rm nn}
\langle d^{\dag}_{\mathbf{n}+\mathbf{n}_{\ell}i{\cal A}\sigma}d^{\phantom{\dag}}_{\mathbf{n}i{\cal B}\bar{\sigma}}\rangle\,,
\end{equation}
where
$\mathbf{n}_{1}=(0,\,0)$,
$\mathbf{n}_{2}=(1,\,0)$,
$\mathbf{n}_{3}=(0,\,1)$,
$\bar{\sigma}=-\sigma$,
and
$V_{\rm nn}=V (|\bm{\delta}|)$
is the in-plane nearest-neighbor Coulomb repulsion. We take
$V_{\text{nn}}/U=0.59$,
in agreement with
Ref.~\onlinecite{Wehling}.

Finally, we introduce the interlayer SDW order parameter
\begin{equation}
\label{Mmn}
B^{rs}_{\mathbf{m};\mathbf{n}\sigma}=V(\mathbf{r}^{1r}_{\mathbf{m}}-\mathbf{r}^{2s}_{\mathbf{n}})
\langle d^{\dag}_{\mathbf{m}1r\sigma}d^{\phantom{\dag}}_{\mathbf{n}2s\bar{\sigma}}\rangle\,.
\end{equation}
For calculations it is assumed that
$B^{rs}_{\mathbf{m};\mathbf{n}\sigma}$
is non-zero only when sites
$\mathbf{r}_{\mathbf{m}}^{1r}$
and
$\mathbf{r}_{\mathbf{n}}^{2s}$
are sufficiently close. Namely, if the hopping amplitude connecting
$\mathbf{r}_{\mathbf{m}}^{1r}$
and
$\mathbf{r}_{\mathbf{n}}^{2s}$
vanishes in our computation scheme, then, the parameter
$B^{rs}_{\mathbf{m};\mathbf{n}\sigma}$
is also zero. Naturally, the number of non-zero
$B^{rs}_{\mathbf{m};\mathbf{n}\sigma}$
depends on the type of the hopping amplitude parametrization. For
parametrization chosen we have up to three non-zero
$B^{rs}_{\mathbf{m};\mathbf{n}\sigma}$
for a given $\mathbf{n}$, $r$, $s$, and $\sigma$. Assuming that the
screening is small at short distances, we chose the function
$V(\mathbf{r})$
in Eq.~\eqref{Mmn} as
$V(\mathbf{r})\propto1/|\mathbf{r}|$
with
$V(d)=V_{\text{nn}}|\bm{\delta}|/d=0.25U$.
All three types of SDW order parameters are restricted to obey the
superlattice periodicity.

Using these order parameters, the full MAtBLG Hamiltonian can be
approximated by a mean field Hamiltonian, the latter being quadratic in
fermionic operators. The mean field Hamiltonian is uniquely specified by a
$8N_{\rm sc} \times 8N_{\rm sc}$
matrix. This matrix diagonalization allows one to determine the
eigenfunctions
$\Phi_{\mathbf{pG}is\sigma}^{(S)}$
and eigenvalues
$E^{(S)}_{\bf p}$
of the mean field Hamiltonian, as well as the mean field ground state
energy for a fixed $n$. The Bogolyubov transformation
\begin{equation}
\label{psi_definition}
d^{\phantom{\dag}}_{\mathbf{pG}is\sigma}
=
\sum_{S}\Phi_{\mathbf{pG}is\sigma}^{(S)}\psi_{\mathbf{p}S}\,,
\end{equation}
introduces new Fermi operators
$\psi_{\mathbf{p}S}$
that diagonalize the mean field Hamiltonian.

Minimizing the mean field ground state energy, one derives the
self-consistency equations for
$\Delta_{\mathbf{n}is}$,
$A^{(\ell)}_{\mathbf{n}i\sigma}$,
and
$B^{rs}_{\mathbf{m};\mathbf{n}\sigma}$.
These equations must be solved numerically for different values of doping
$n$ confined to the interval
$-4 < n <4$.
The details of the numerical procedure can be found in
Ref.~\onlinecite{Nematic}.

\subsection{Symmetry properties of the order parameters}

The results of the order parameters calculations for different doping
levels are presented in
Ref.~\onlinecite{Nematic},
where spatial profiles of
$\Delta_{\mathbf{n}is}$
and
$A^{(\ell)}_{\mathbf{n}i\sigma}$
are plotted. Let us briefly describe their main properties. The order
parameters are non-zero within the doping range
$-4<n<4$.
The absolute values of the order parameters decrease to zero when
$|n| \rightarrow 4$.
For any doping, the absolute values of
$A^{(\ell)}_{\mathbf{n}i\sigma}$
are smaller than
$\Delta_{\mathbf{n}is}$,
and the values of
$B^{rs}_{\mathbf{m};\mathbf{n}\sigma}$
are by order of magnitude smaller than
$A^{(\ell)}_{\mathbf{n}i\sigma}$.
All three types of the order parameters have maximum values inside the AA
region of the superlattice cell because electrons at the Fermi level are
located mainly in this region.

The order parameter
$\Delta_{\mathbf{n}is}$
describes the on-site spins polarized in the $xy$ plane. At zero doping,
$\Delta_{\mathbf{n}is}$
can be chosen to be real for all
$\mathbf{n}$,
$i$, and $s$, that is, all spins are collinear and parallel or antiparallel
to the $x$ axis. Our simulations show that
$\Delta_{\mathbf{n}i{\cal A}}=-\Delta_{\mathbf{n}i{\cal B}}$
with a good accuracy. Thus, we have an antiferromagnetic ordering of spins.
At finite doping, the on-site spins are no longer collinear, but they
remain coplanar. In this case, we observe a kind of helical
antiferromagnetic ordering. Note that in present simulations we do not
allow on-site spins to have the $z$ component. However, similar
calculations performed in
Ref.~\onlinecite{OurJETPLetters2022}
showed that the coplanar spin texture survives even if we allow for
spin non-coplanarity.

As for on-link order parameter
$A^{(\ell)}_{\mathbf{n}i\sigma}$,
at zero doping these vectors are collinear, while at finite doping they are
coplanar. Similar to the on-site spins, simulations performed in
Ref.~\onlinecite{OurJETPLetters2022}
showed that
$A^{(\ell)}_{\mathbf{n}i\sigma}$
remain coplanar (with the exception of several on-link spins, see Fig.~3 of Ref.~\onlinecite{OurJETPLetters2022}) even if non-coplanarity is permitted by the minimization
algorithm.

An important observation for the present study is that the doping reduces
the symmetry of the order parameters. They have the hexagonal symmetry at
zero doping, which is the symmetry of the crystal. Specifically, the order
parameters are invariant under rotation on
$60^{\circ}$
around the center of the AA region. Doping reduces the symmetry from
$C_6$
to
$C_2$.
For example, near the half-filling, the order parameters are invariant under rotation on
$180^{\circ}$
around the center of the AA region. Reduction of the symmetry of the order
parameters affects the symmetry of the mean-field spectrum, indicating the
appearance of a electron nematic state under doping. At zero doping the
mean field spectrum has the hexagonal symmetry. At finite doping the
symmetry of the spectrum is reduced; the eigenenergies
$E_{\mathbf{p}}^{(S)}$
are invariant under rotation of vector
$\mathbf{p}$
on
$180^{\circ}$
(but not on
$60^{\circ}$)
around the
$\bm{\Gamma}$
point.

The reduced symmetry of the order parameters affects the symmetry of the
local density of states (see Fig.~6. of Ref.~\onlinecite{Nematic}).
``Nematic" features of the local density of states were detected in
STM measurements in
Refs.~\onlinecite{MottNematicNature2019,KerelskyNematicNature2019}.
In these experiments, the bright spots in STM images, centered at the AA
regions of the moir{\'{e}} superlattice, were uniaxi-ally stretched.

\section{Polarization operator and screened Coulomb interaction}
\label{PolarizationOperator}

In our simulations, the SDW order parameter is a short-range one: it
includes on-site and nearest-neighbor terms. At small distances, that is,
at large momenta
$k\sim1/a$
the system behaves as two decoupled graphene layers. In such a limit, the
screening does not introduce new qualitative features. Indeed, the static
polarization operator of the graphene layer
equals~\cite{InteractionsGrapheneReview2012}
$\Pi(q)=-q/(4v_{\rm F})$,
where
$v_{\rm F}=\sqrt{3}ta/2$
is the Fermi velocity of the graphene. As a result, the effective Coulomb
interaction can be estimated as
\begin{equation}
\label{Veff0}
V(q)=\frac{1}{\epsilon_{\rm RPA}}\frac{2\pi e^2}{q}\,,
\end{equation}
where the dielectric constant of the bilayer is
$\epsilon_{\rm RPA}=\epsilon+\pi e^2/v_{\rm F}$
($\epsilon$ is the dielectric constant of the media surrounding the
sample). According to this formula, in the real space representation we
have
$V(r)\propto1/r$,
and the interaction slowly decays with the distance. This is why we used
$1/r$
dependence to estimate the interlayer interaction in constructing our
short-range order parameter.

Such arguments are not applicable to a superconducting phase since the
stabilization of the superconducting order parameter relies on the
interaction with small transferred momenta. At large distances and small
momenta,
Eq.~\eqref{Veff0}
fails for MAtBLG, and the peculiarities of the system, such as moir{\'{e}}
structure, the flat-bands formation, the SDW order, must be accounted for.
We do this in the RPA approximation, using the wave functions and the
eigenenergies corresponding to the SDW mean-field Hamiltonian.

To find the RPA interaction, the polarization operator has to be calculated
first. It is a matrix function of the transferred momentum
$\mathbf{q}$
defined
as~\cite{Triola2012}
\begin{eqnarray}
\label{PG}
\Pi_{\mathbf{Q}\mathbf{Q}'}^{isjs'}(\mathbf{q})
&=&
\frac{1}{N_{sc}}\sum_{SS'}
	\int\!\frac{d^2\mathbf{p}}{v_{\rm RBZ}}
		\frac{
			n_{\rm F}(E_{\mathbf{p}}^{(S)})
			-
			n_{\rm F}(E_{\mathbf{p}+\mathbf{q}}^{(S')})
		}
		{E_{\mathbf{p}}^{(S)}-E_{\mathbf{p}+\mathbf{q}}^{(S')}}
\nonumber
\\
&&\times\left( \sum_{\mathbf{G}\sigma}
	\Phi_{\mathbf{p}\mathbf{G}is\sigma}^{(S)}
	\Phi_{\mathbf{p}+\mathbf{q}\mathbf{G}+\mathbf{Q}is\sigma}^{(S')*}
\right)
\nonumber
\\
&&\times\left(\sum_{\mathbf{G}'\sigma'}
	\Phi_{\mathbf{p}\mathbf{G}'js'\sigma'}^{(S)*}
	\Phi_{\mathbf{p}+\mathbf{q}\mathbf{G}'+\mathbf{Q}'js'\sigma'}^{(S')}
\right),
\end{eqnarray}
where
$v_{\rm RBZ}$
is the Brillouin zone area of the superlattice, and
$n_{\rm F}(E)$
is the Fermi function. In
Eq.~\eqref{PG}
the reciprocal superlattice vector
$\mathbf{Q}$
(vector
$\mathbf{Q}'$)
is confined to the first Brillouin zone of the layer~$i$ (layer~$j$).
The momentum integration is performed over the reduced
Brillouin zone.

Within RPA, the renormalized interaction
$\hat{V}$
satisfies the equation
\begin{equation}
\label{VRPA}
\hat{V}=\hat{V}^{(0)}-\hat{V}^{(0)}\hat{\Pi}\hat{V}\,.
\end{equation}
Here the matrix-valued function representing the bare interaction
$\hat{V}^{(0)}(\mathbf{q})=V_{\mathbf{Q}\mathbf{Q}'}^{(0)isjs'}$
can be written as
\begin{eqnarray}
\label{summation}
V_{\mathbf{Q}\mathbf{Q}'}^{(0)isjs'}(\mathbf{q})
&=&
\frac{1}{N_{sc}}
\sum_{\mathbf{nm}}
	e^{-i(\mathbf{q}+\mathbf{Q})
		(\mathbf{r}_{\mathbf{n}}^{is}
		-
		\mathbf{r}_{\mathbf{m}}^{js'})
	}
	\times
\nonumber\\
&&V(\mathbf{r}_{\mathbf{n}}^{is}-\mathbf{r}_{\mathbf{m}}^{js'})
	e^{i(\mathbf{Q}'-\mathbf{Q})\mathbf{r}_{\mathbf{m}}^{js'}}.
\end{eqnarray}
where
${\bf r}_\mathbf{m}^{j s'}$
runs over the atoms located inside zeroth superlattice cell, while
${\bf r}_\mathbf{n}^{i s}$
runs over all atoms of the sample.

In Eq.~\eqref{summation} we neglect the Hubbard term. In separate simulations we
showed that adding the Hubbard term does not change the results
significantly. This is because at small transferred momenta (the case,
which is of interest for us in the part concerning the superconductivity)
the intersite Coulomb term dominates.
This is not surprising as the
screening ultimately fails at short distances.

\begin{figure}[t]
\centering
\includegraphics[width=0.99\columnwidth]{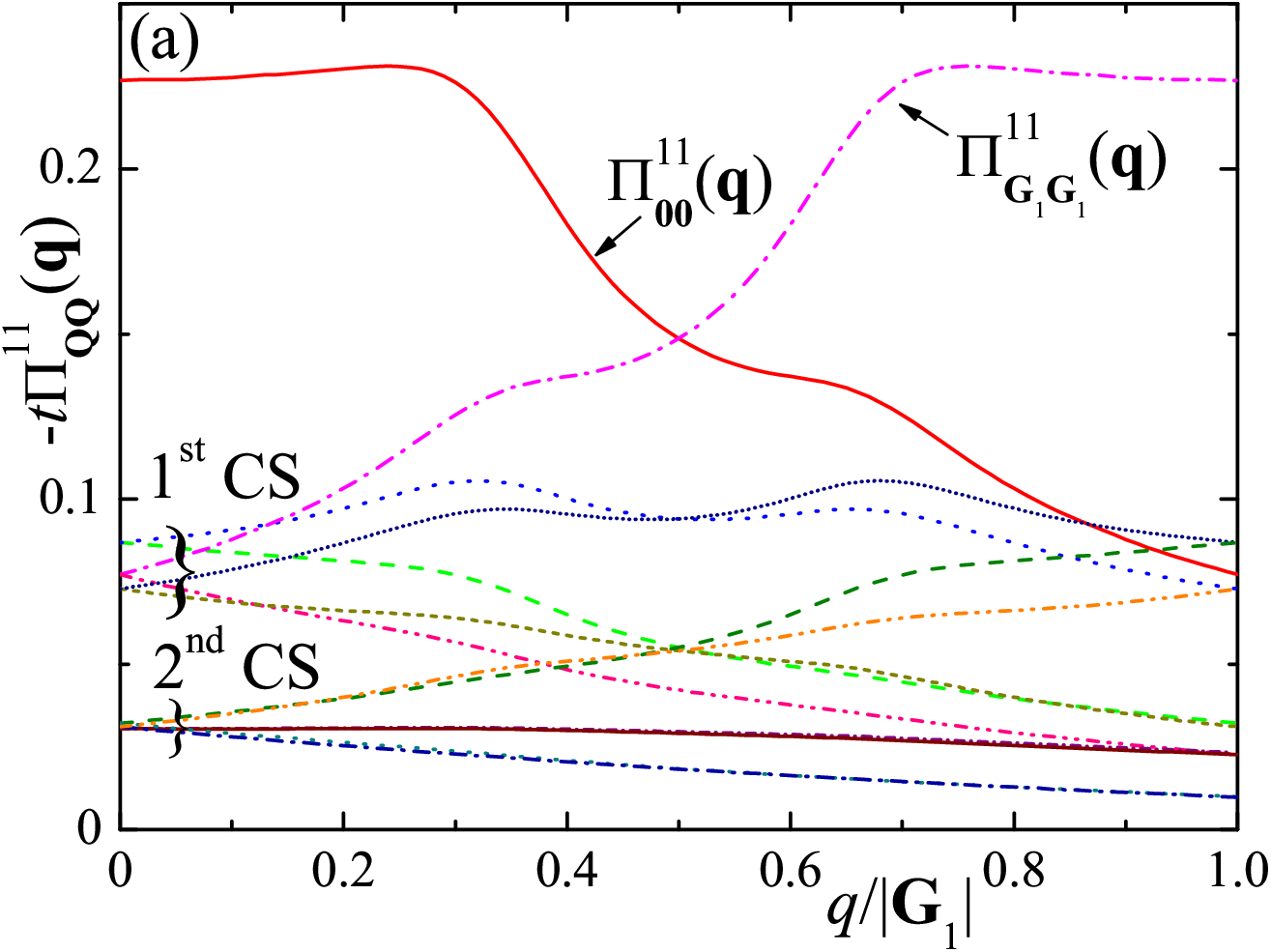}\\
\includegraphics[width=0.99\columnwidth]{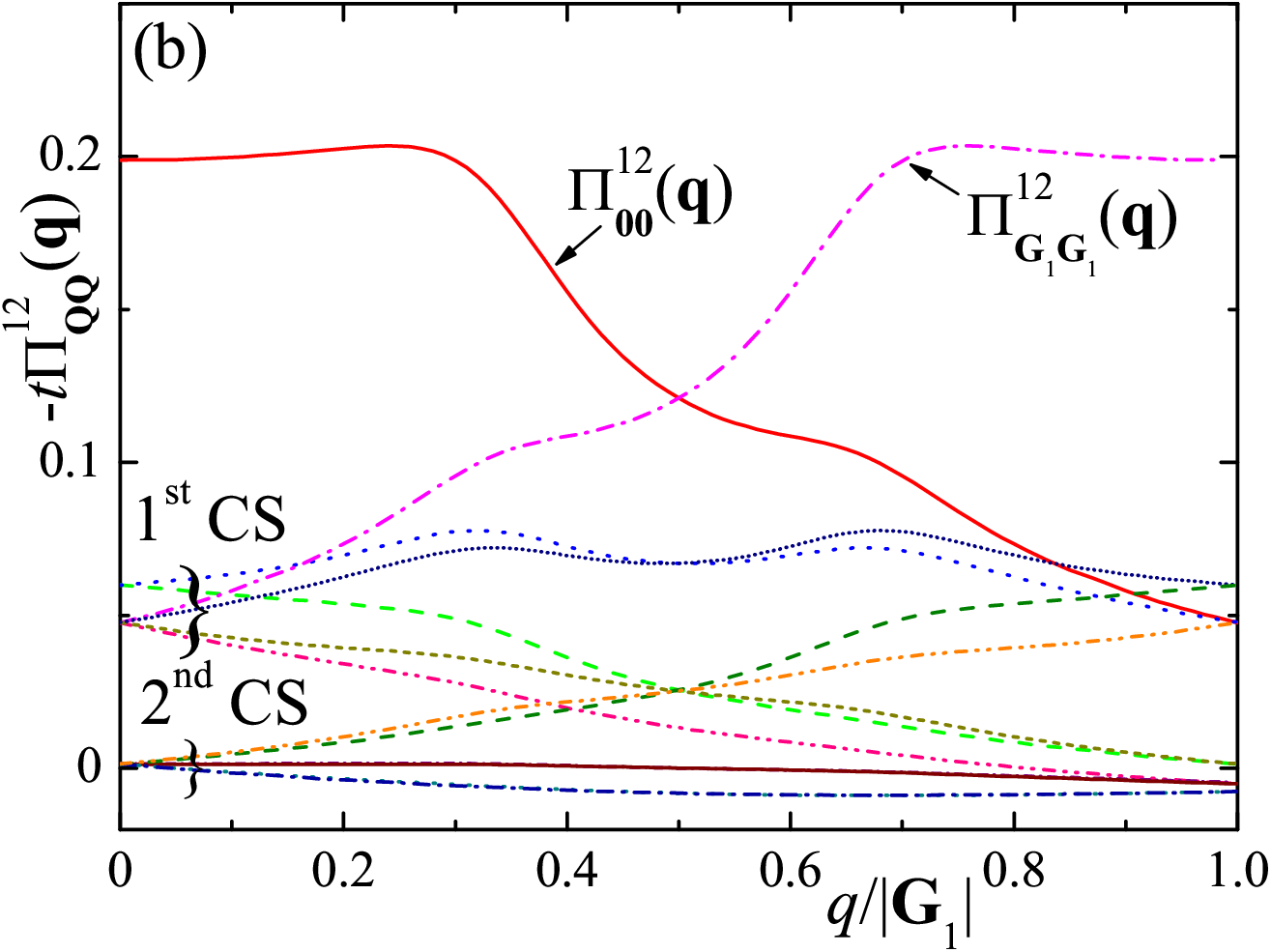}
\caption{The components of the polarization operator
$\Pi_{\mathbf{Q}\mathbf{Q}}^{11}(\mathbf{q})$
(a) and
$\Pi_{\mathbf{Q}\mathbf{Q}}^{12}(\mathbf{q})$
(b) calculated in the SDW phase at
$\mathbf{q}=q\bm{{\cal G}}_1/|\bm{{\cal G}}_1|$
for
$\mathbf{Q}=0$
and
$\mathbf{Q}$
belonging to the first two coordination spheres. The doping level is
$n=-1.75$.
\label{FigP}
}
\end{figure}

At small
$\mathbf{q}+\mathbf{Q}$
one can obtain an analytical expression for the matrix
$V_{\mathbf{Q}\mathbf{Q}'}^{(0)isjs'}$.
In the case
$i=j$,
the translation symmetry allows us to convert the summation over
${\bf r}_\mathbf{n}^{is}$
to the summation over
${\bf r}_\mathbf{n}^{is} - {\bf r}_\mathbf{m}^{is'}$,
and the summation over
$\mathbf{m}$
gives a factor
$N_{sc}\delta_{\mathbf{QQ}'}$
before the sum in
Eq.~\eqref{summation}.
Further, when
$\mathbf{q}+\mathbf{Q}$
is small
($|{\bf q} + {\bf Q}| \ll a^{-1}$),
the lattice summation can be replaced by the space integration. As a
result, we establish
\begin{equation}
\label{V0ii}
V_{\mathbf{Q}\mathbf{Q}'}^{(0)1s1s'}(\mathbf{q})
=
V_{\mathbf{Q}\mathbf{Q}'}^{(0)2s2s'}(\mathbf{q})
=
\delta_{\mathbf{QQ}'}\frac{2\pi e^2}{\epsilon v_c|\mathbf{q}+\mathbf{Q}|},
\end{equation}
where
$v_c=\sqrt{3}a^2/2$
is the area of the graphene unit cell. For
$i\neq j$,
one can find such
$\mathbf{m}$
that
$\mathbf{r}_{\mathbf{m}}^{2s'}
=
d\mathbf{e}_z+\mathbf{r}_{\mathbf{n}}^{1s}+\bm{\delta}_{\mathbf{nm}}^{ss'}$,
where
$\bm{\delta}_{\mathbf{nm}}^{ss'}$
is small
($|\bm{\delta}_{\mathbf{nm}}^{ss'}| \lesssim a$).
At small
$|{\bf q} + {\bf Q}|$
we can neglect
$\bm{\delta}_{\mathbf{nm}}^{ss'}$
and replace the summation by integration. This allows us to derive
\begin{equation}
\label{V0ij}
V_{\mathbf{Q}\mathbf{Q}'}^{(0)1s2s'}(\mathbf{q})
=
V_{\mathbf{Q}\mathbf{Q}'}^{(0)2s1s'}(\mathbf{q})
=
\delta_{\mathbf{QQ}'}
\frac{2\pi e^2e^{-|\mathbf{q}+\mathbf{Q}|d}}
{\epsilon v_c|\mathbf{q}+\mathbf{Q}|}.
\end{equation}
\begin{figure}[t]
\centering
\includegraphics[width=0.99\columnwidth]{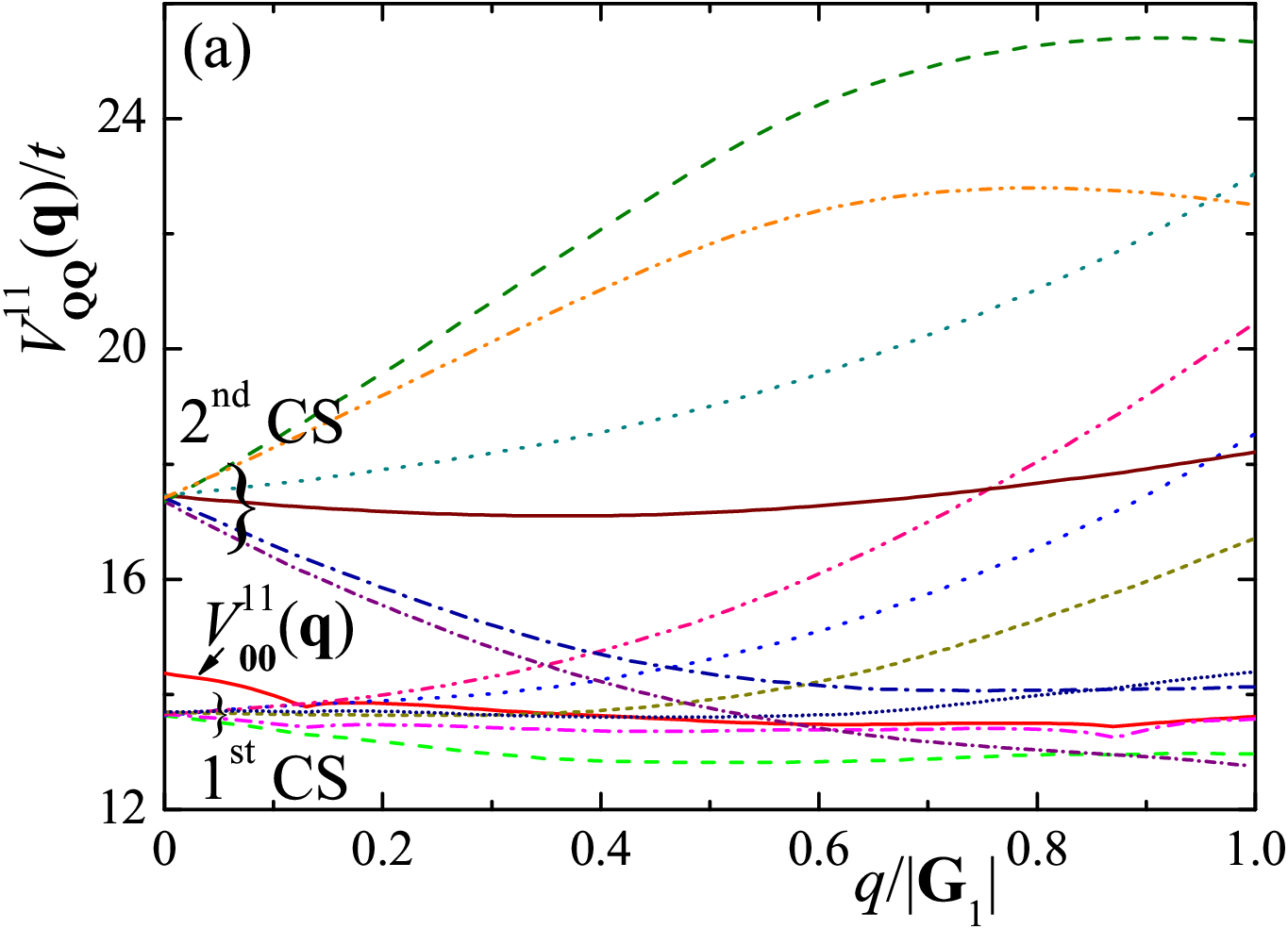}\\
\includegraphics[width=0.99\columnwidth]{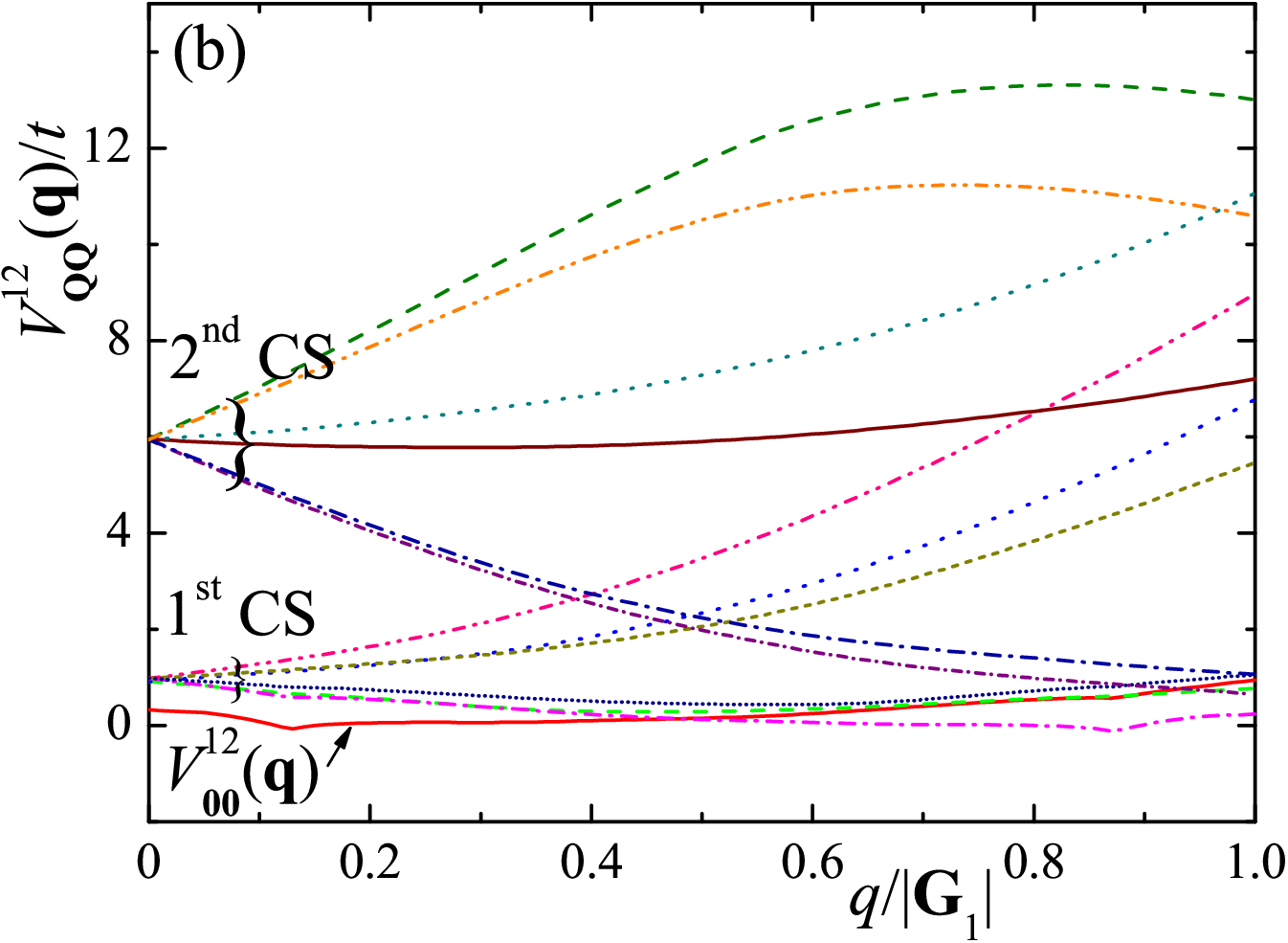}
\caption{The dependence of renormalized Coulomb interaction
$V_{\mathbf{Q}\mathbf{Q}}^{11}(\mathbf{q})$
(a) and
$V_{\mathbf{Q}\mathbf{Q}}^{12}(\mathbf{q})$
(b) calculated in the SDW phase at
$\mathbf{q}=q\bm{{\cal G}}_1/|\bm{{\cal G}}_1|$
for
$\mathbf{Q}=0$
and for
$\mathbf{Q}$
belonging to the first two coordination spheres. The doping level is
$n=-1.75$
and
$\epsilon=1$.
\label{FigV}
}
\end{figure}
In our simulations, we use truncated matrices
$\Pi_{\mathbf{Q}\mathbf{Q}'}^{isjs'}$
and
$V_{\mathbf{Q}\mathbf{Q}'}^{(0)isjs'}$
with
$\mathbf{Q}$
and
$\mathbf{Q}'$
being restricted to the insides of the 11th coordination sphere (CS). The
total number of such
$\mathbf{Q}$
is $91$.

According to the
expressions~\eqref{V0ii}
and~\eqref{V0ij},
the bare interaction
$V_{\mathbf{Q}\mathbf{Q}'}^{(0)isjs'}$
at small transferred momentum is independent of the sublattice indices $s$,
$s'$. Then, one can prove that the screened interaction is also independent
of these indices
$V_{\mathbf{Q}\mathbf{Q}'}^{isjs'} = V_{\mathbf{Q}\mathbf{Q}'}^{ij}$.
The matrix
$V_{\mathbf{Q}\mathbf{Q}'}^{ij}$
satisfies
Eq.~\eqref{VRPA}
with the sublattice-independent polarization operator
$\Pi_{\mathbf{Q}\mathbf{Q}'}^{ij}$
defined as
\begin{eqnarray}
\label{PGij}
\Pi_{\mathbf{Q}\mathbf{Q}'}^{ij}
=
\sum_{ss'}\Pi_{\mathbf{Q}\mathbf{Q}'}^{isjs'}.
\end{eqnarray}

We calculate the polarization operator numerically for different doping
levels. The temperature is chosen as
$T=5\times10^{-3}W_{\text{SDW}}$,
where
$W_{\text{SDW}}$
is the width of the eight low-energy mean-field bands. It is not possible
to perform the double summation in
Eq.~\eqref{PG}
over all bands at realistic time. For this reason we keep only $104$ bands
closest to the Fermi level in the summation over $S$ and $S'$, assuming
that the contributions from higher energy bands are small. The functions
$\Pi_{\mathbf{Q}\mathbf{Q}}^{11}(\mathbf{q})$
and
$\Pi_{\mathbf{Q}\mathbf{Q}}^{12}(\mathbf{q})$
are shown in
Fig.~\ref{FigP}
for
$\mathbf{Q}=0$
and for
$\mathbf{Q}$
belonging to the first two CS. These results correspond to the doping level
$n=-1.75$.
The vector
$\mathbf{q}$
is along the vector
$\bm{{\cal G}}_1$.
We see that
$-\Pi_{\mathbf{00}}^{ij}(\mathbf{q})$
decreases with
$\mathbf{q}$.
The values of
$-\Pi_{\mathbf{Q}\mathbf{Q}}^{ij}(\mathbf{0})$
decrease with the increase of the absolute values of
$\mathbf{Q}$.
Our simulations show that
$\Pi_{\mathbf{Q}\mathbf{Q}'}^{ij}(\mathbf{q})$
is almost diagonal in
$\mathbf{Q}$
and
$\mathbf{Q}'$
for
$\mathbf{Q}$
or
$\mathbf{Q}'$
belonging to the third and larger CS. In this case we have
$\Pi_{\mathbf{Q}\mathbf{Q}'}^{ij}(\mathbf{q})
\cong
\delta_{\mathbf{Q}\mathbf{Q}'}\Pi_{\mathbf{Q}}^{ij}(\mathbf{q})$.

\begin{figure}[t]
\centering
\includegraphics[width=0.99\columnwidth]{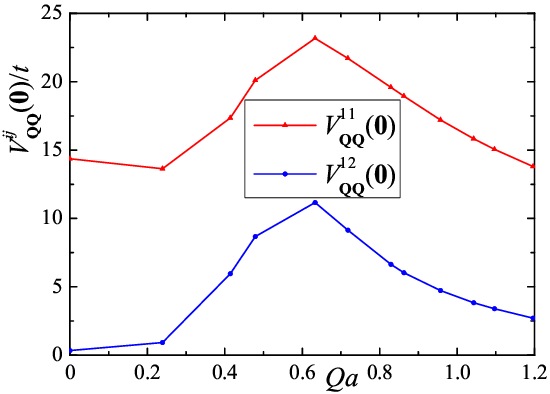}
\caption{The dependence of
$V_{\mathbf{Q}\mathbf{Q}}^{11}(\mathbf{0})$
(red triangles) and
$V_{\mathbf{Q}\mathbf{Q}}^{12}(\mathbf{0})$
(blue circles) on $Q$. The doping level is
$n=-1.75$
and
$\epsilon=1$.
\label{FigV0}
}
\end{figure}

\begin{figure*}[t]
\centering
\includegraphics[width=0.32\textwidth]{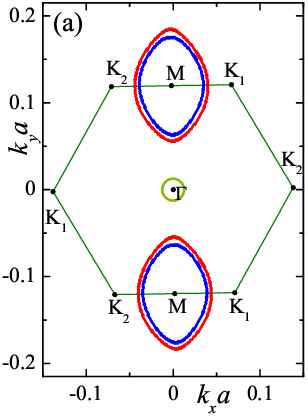}
\includegraphics[width=0.32\textwidth]{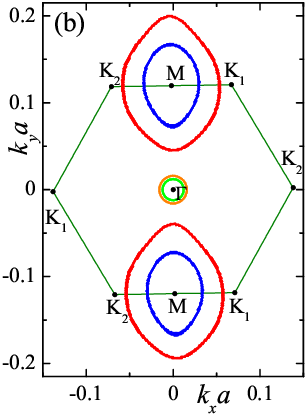}
\includegraphics[width=0.32\textwidth]{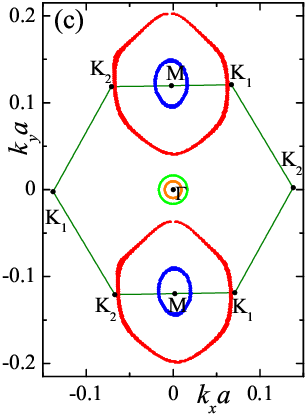}
\caption{The Fermi surface in the SDW phase calculated for
$n=-1.75$~(a),
$n=-1.69$~(b),
and
$n=-1.67$~(c).
For each doping there are two elliptical Fermi surface sheets centered at
$\mathbf{M}$
point and two circular Fermi surface sheets centered at
$\bm{\Gamma}$
point. In each plot the upper and lower
$\mathbf{M}$
points are equivalent. Due to nematicity of the underlying
spin texture, the Fermi surface symmetry is $C_2$, not $C_6$.
\label{FigFS}
}
\end{figure*}

Figure~\ref{FigV}
shows the dependence of the renormalized Coulomb interaction
$V_{\mathbf{Q}\mathbf{Q}}^{11}(\mathbf{q})$
and
$V_{\mathbf{Q}\mathbf{Q}}^{12}(\mathbf{q})$
on $q$ calculated for
$\mathbf{Q}=0$
and for
$\mathbf{Q}$
belonging to the first two CS. We see, that
$V_{\mathbf{Q}\mathbf{Q}}^{ij}(\mathbf{0})$
increases with the increase of $Q$. Such a dependence exists up to the
fourth CS. At larger $Q$, the
$V_{\mathbf{Q}\mathbf{Q}}^{ij}(\mathbf{0})$
decreases approximately as
$1/Q$.
The dependence of
$V_{\mathbf{Q}\mathbf{Q}}^{ij}(\mathbf{0})$
on $Q$ is shown in
Fig.~\ref{FigV0}.

Observe that according to the results presented in
Figs.~\ref{FigV}
and~\ref{FigV0}
the interlayer interaction turns out to be noticeably weaker than the
intra-layer one. This is because the factor
$e^{-qd}$
appearing in the
definition~\eqref{V0ij}
of the bare interlayer interaction cannot be neglected even at small
transfer momentum. More information about the difference between
intra-layer and interlayer interactions in bilayer graphene systems can be
found in
Refs.~\onlinecite{SuperconductivityAB,SuperconductivityAA}.

\section{Superconductivity}
\label{superconductivity}

We examine a possibility of the superconducting state controlled by the
renormalized Coulomb interaction near the half-filling, where it was
observed
experimentally~\cite{NatureSC2018,MottSCNature2019}.
For each momentum
$\mathbf{p}$
in the reduced Brillouin zone we arrange energies of the low-energy bands
$E_{\mathbf{p}}^{({\cal S})}$
(${\cal S}=1,\,2,\dots,\,8$)
in ascending order. In our study of the superconductivity we consider three
doping levels:
$n=-1.75$,
$n=-1.69$,
and
$n=-1.67$.

In our scenario the superconductivity becomes possible since the SDW order
cannot completely eliminate Fermi surface of MAtBLG. Thus, the remaining
low-lying fermionic degrees of freedom can become unstable in the
superconductivity channel. The Fermi surface structures corresponding to
the three doping levels are shown in
Fig.~\ref{FigFS}.
For each doping there are two almost elliptical Fermi surface sheets
centered at
$\mathbf{M}$
point and two circular Fermi surface sheets centered at
$\bm{\Gamma}$
point. Elliptical Fermi surfaces are formed by the bands with
${\cal S}_1=3$
(bigger ellipse) and
${\cal S}_2=4$
(smaller ellipse), while circular Fermi surface sheets are formed by the
bands with
${\cal S}_3=1$
and
${\cal S}_4=2$.
For
$n=-1.75$
the sizes of ellipses are almost equal to each other. When we increase the
doping the sizes of the ellipses become more dissimilar. This happens
because the low-energy spectra are almost doubly degenerate at
half-filling, while the bands tend to separate from each other when $n$
approaches $-1$. Note that for the considered doping levels the mean-field
spectra demonstrate nematicity, that is, the spectra have
$C_2$
symmetry group, which is lower than the
$C_6$
symmetry of the crystal. The nematic SDW order induces the nematicity of
the Fermi surface, the latter is clearly visible in
Fig.~\ref{FigFS}.

The bands
${\cal S}_3$
and
${\cal S}_4$
forming the circular Fermi surfaces are not interesting for the
superconducting pairing since they have large Fermi velocities at the Fermi
level and small Fermi momenta. The bands
${\cal S}_1$
and
${\cal S}_2$
forming the elliptical Fermi surfaces around
$\mathbf{M}$
point are more relevant for the superconductivity since their Fermi
velocities are small enough (the density of states is large) and the Fermi
momenta are larger than that for the circular Fermi surfaces.

Using fermionic operators
$\psi_{{\bf p} S}$
introduced in
Eq.~(\ref{psi_definition})
and keeping only terms relevant for the superconducting pairing, one can
rewrite the renormalized interaction Hamiltonian as follows
\begin{equation}
\label{HintSC}
H_{\text{int}}=\frac{1}{2{\cal N}}\sum_{\mathbf{pp}'}\sum_{\cal{SS}'}\Gamma^{({\cal S},{\cal S}')}_{\mathbf{pp}'}
\psi^{\dag}_{-\mathbf{p}'{\cal S}'}\psi^{\dag}_{\mathbf{p}'{\cal S}'}
\psi^{\phantom{\dag}}_{\mathbf{p}{\cal S}}\psi^{\phantom{\dag}}_{-\mathbf{p}{\cal S}}\,.
\end{equation}
Here and below the summation over
${\cal S}$
and
${\cal S}'$
is performed over bands
${\cal S}_1$
and
${\cal S}_2$
and
\begin{eqnarray}
\Gamma^{({\cal S},{\cal S}')}_{\mathbf{pp}'}
&=&\!\!\!
\sum_{ij\atop\mathbf{Q}_1\mathbf{Q_2}}\!\!
	\left( \sum_{\mathbf{G}s\sigma}
		\Phi_{\mathbf{pG}is\sigma}^{({\cal S})}
		\Phi_{\mathbf{p}'\mathbf{G}+\mathbf{Q}_1is\sigma}^{({\cal S}')*}
	\right)
	V^{ij}_{\mathbf{Q}_1\mathbf{Q}_2}(\mathbf{p}'-\mathbf{p})
\nonumber
\\
&\times&\!\!\!
\left(\sum_{\mathbf{G}s\sigma}
	\Phi_{-\mathbf{pG}js\sigma}^{({\cal S})}
	\Phi_{-\mathbf{p}'\mathbf{G}-\mathbf{Q}_2js\sigma}^{({\cal S}')*}
\right)
\end{eqnarray}
is the effective interaction in the Cooper channel.

We assume that in the superconducting state the following expectation
values are non-zero
\begin{equation}
\label{eta}
\alpha^{({\cal S})}_{\mathbf{p}}=\left\langle\psi^{\dag}_{-\mathbf{p}{\cal S}}\psi^{\dag}_{\mathbf{p}{\cal S}}\right\rangle\,.
\end{equation}
The total momentum of the pair is zero. We introduce the superconducting
order parameter in the form
\begin{equation}
\label{DeltaSC}
\Delta^{({\cal S})}_{\mathbf{p}}=\frac{1}{\cal N}\sum_{{\cal S}'\mathbf{p}'}\Gamma^{({\cal S},{\cal S}')}_{\mathbf{pp}'}\alpha^{({\cal S}')}_{\mathbf{p}'}\,.
\end{equation}
Transforming the interaction
Hamiltonian~\eqref{HintSC}
to its mean-field version, we derive the self-consistency equation for the
order parameter. After standard calculations we obtain
\begin{eqnarray}
\label{GapSCeq}
\Delta^{({\cal S})}_{\mathbf{p}}
&=&
-\sum_{{\cal S}'}
	\int\frac{d^2\mathbf{p}'}{v_{\rm BZ}}
		\frac{
			\Gamma^{({\cal S},{\cal S}')}_{\mathbf{pp}'}
			\Delta^{({\cal S}')}_{\mathbf{p}'}
		}
		{2\sqrt{\left[E^{({\cal S}')}_{\mathbf{p}'}-\mu\right]^2
			+
			\left|\Delta^{({\cal S}')}_{\mathbf{p}'}\right|^2}
		}
\nonumber
\\
&\times&
\tanh\left[
	\frac{1}{2T}
	\sqrt{
		\left[E^{({\cal S}')}_{\mathbf{p}'}-\mu\right]^2
		+
		\left|\Delta^{({\cal S}')}_{\mathbf{p}'}\right|^2
	}
\right]\,,
\end{eqnarray}
where $v_{\rm BZ}$ is the Brillouin zone area of the graphene and the integration is performed over the reduced Brillouin zone.
\begin{figure}[t]
\centering
\includegraphics[width=0.99\columnwidth]{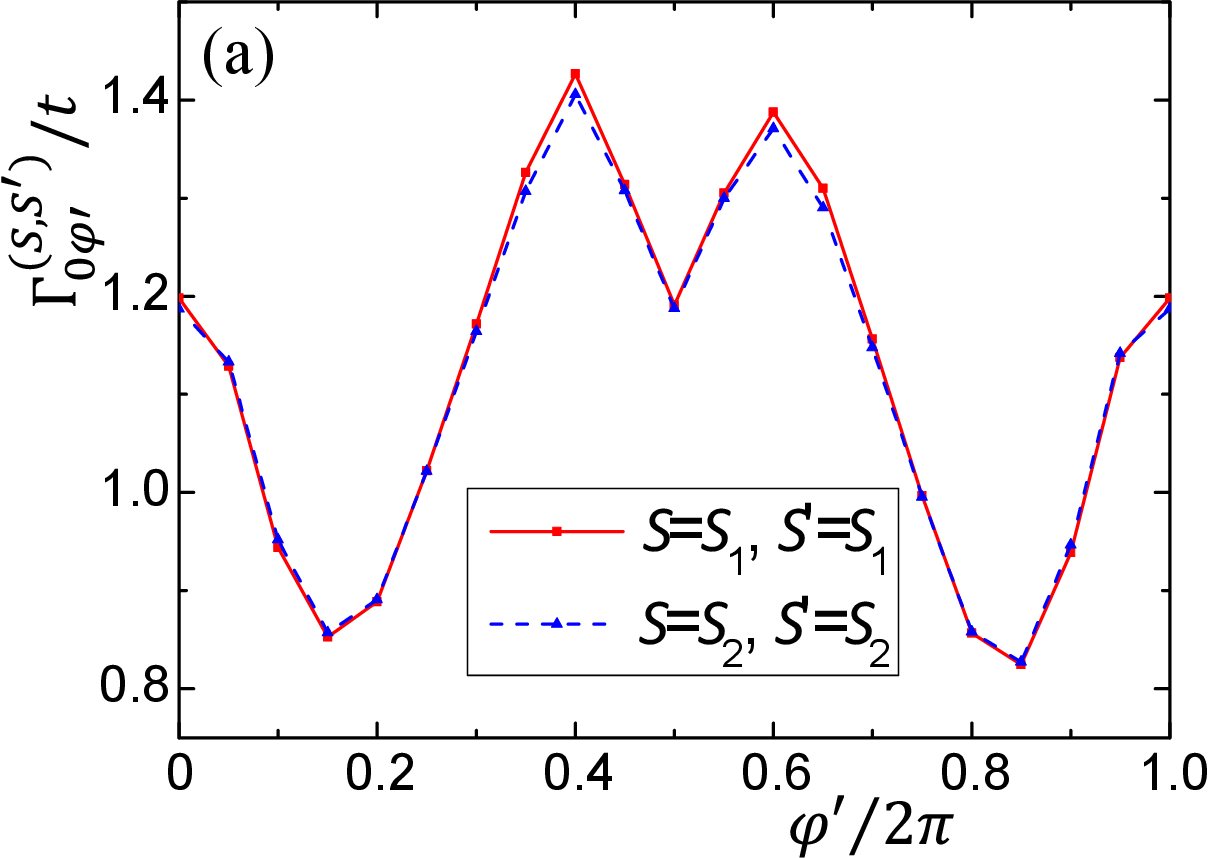}\\
\includegraphics[width=0.99\columnwidth]{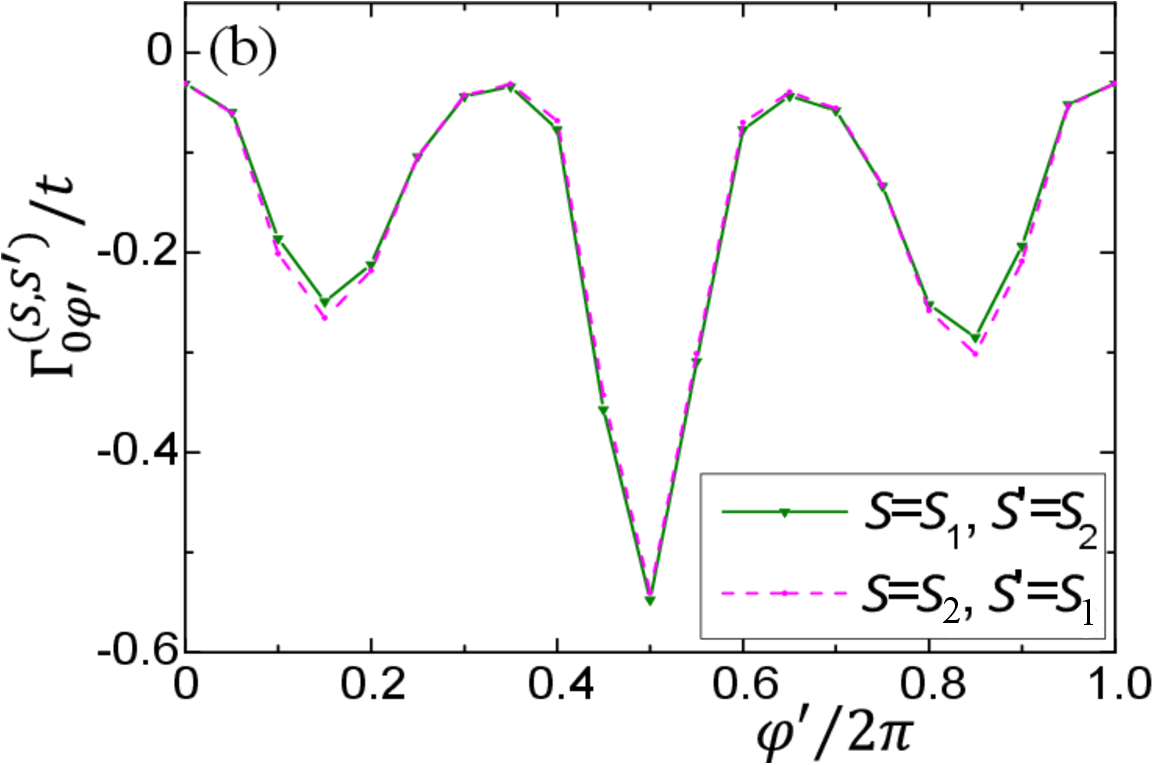}
\caption{Interaction in the Cooper channel. The dependence of
$\tilde{\Gamma}^{({\cal S},{\cal S}')}_{0\varphi'}$
on $\varphi'$, calculated for
$n=-1.75$.
Panel (a) corresponds to
${\cal S}={\cal S}_1,\,{\cal S}'={\cal S}_1$
and
${\cal S}={\cal S}_2,\,{\cal S}'={\cal S}_2$,
while panel (b) corresponds to
${\cal S}={\cal S}_1,\,{\cal S}'={\cal S}_2$
and
${\cal S}={\cal S}_2,\,{\cal S}'={\cal S}_1$,
see the legend.
\label{FigGamma}
}
\end{figure}

We do not solve the integral
equation~\eqref{GapSCeq},
but only estimate the critical temperature
$T_c$
by order of magnitude. With a good accuracy the Fermi surface sheets
centered at
$\mathbf{M}$
point have shapes of ellipses. One can introduce the polar angle $\varphi$
and parameterize the Fermi momenta of the bands
${\cal S}_1$
and
${\cal S}_2$
as
$\mathbf{p}_{\rm F}^{({\cal S})}(\varphi)
=
\mathbf{M}+\mathbf{k}_{\rm F}^{({\cal S})}(\varphi)$,
where
\begin{equation}
\label{kFphi}
\mathbf{k}_{\rm F}^{({\cal S})}(\varphi)
=
k_{1F}^{({\cal S})}\mathbf{n}_1\cos\varphi
+
k_{2F}^{({\cal S})}\mathbf{n}_2\sin\varphi\,.
\end{equation}
In this equation,
$k_{1F}^{({\cal S})}$
and
$k_{2F}^{({\cal S})}$
are found by fitting of the
${\cal S}$'th
Fermi surface sheet by an ellipse, and
$\mathbf{n}_1$
and
$\mathbf{n}_2$
are the unit vectors parallel and perpendicular to the vector
$\bm{{\cal G}}_2$,
correspondingly. Near the
${\cal S}$'th
Fermi surface sheet, one can write the energy of the band
${\cal S}$
as
\begin{equation}
\label{ESappr}
E_{\mathbf{M}+\mathbf{k}}^{({\cal S})}
\cong
\mu+\mathbf{v}^{({\cal S})}(\varphi) \cdot
\left[
	\mathbf{k}-\mathbf{k}_{\rm F}^{({\cal S})}(\varphi)
\right],
\end{equation}
where
\begin{equation}
\mathbf{v}^{({\cal S})}(\varphi)
=
\left.\frac{\partial E_{\mathbf{M}+\mathbf{k}}^{({\cal S})}}
	{\partial\mathbf{k}}
\right|_{\mathbf{k}
=
\mathbf{k}_{\rm F}^{({\cal S})}(\varphi)}\,.
\end{equation}
In
Eq.~\eqref{GapSCeq}
we introduce for each
${\cal S}'$
the polar coordinates
$(k,\,\varphi)$
in the integral over
$\mathbf{p}'$
as follows
\begin{equation}
\label{ppolar}
\mathbf{p}'
=
\mathbf{M}
+
k\left(
	\mathbf{n}_1\cos\varphi
	+
	\varkappa^{({\cal S})}\mathbf{n}_2\sin\varphi
\right),
\end{equation}
where
$\varkappa^{({\cal S})}=k_{2F}^{({\cal S})}/k_{1F}^{({\cal S})}$.
In this case, we have
$d^2\mathbf{p}'=\varkappa^{({\cal S})}kdkd\varphi$.
Using
Eqs.~\eqref{kFphi}
and~\eqref{ppolar}
we can rewrite
Eq.~\eqref{ESappr}
in the form
\begin{equation}
\label{ESappr2}
E_{\mathbf{M}+\mathbf{k}}^{({\cal S})}\cong\mu+\tilde{v}^{({\cal S})}(\varphi)\left(k-k_{1F}^{({\cal S})}\right),
\end{equation}
where
\begin{equation}
\tilde{v}^{({\cal S})}(\varphi)
=
\mathbf{v}^{({\cal S})}(\varphi)
\cdot
\left(
	\mathbf{n}_1\cos\varphi
	+
	\varkappa^{({\cal S})}\mathbf{n}_2\sin\varphi
\right).
\end{equation}
We replace
$\Gamma^{({\cal S},{\cal S}')}_{\mathbf{pp}'}$
in Eq.~\eqref{GapSCeq} by their values at Fermi momenta introducing the functions
\begin{equation}
\label{Gamm_SS}
\tilde{\Gamma}^{({\cal S},{\cal S}')}_{\varphi\varphi'}
=
\Gamma^{({\cal S},{\cal S}')}_{{\bf p} {\bf p}'}
\Big|_{{{\bf p} = \mathbf{M}+\mathbf{k}_{\rm F}^{({\cal S})}(\varphi) \atop
{\bf p}' = \mathbf{M}+\mathbf{k}_{\rm F}^{({\cal S}')}(\varphi')}}\,.
\end{equation}
Finally, we assume the following ansatz for the superconducting order
parameter
\begin{equation}
\Delta^{({\cal S})}_{\mathbf{M}+\mathbf{k}}
=
\left\{\begin{array}{cl}
	\Delta^{({\cal S})}(\varphi)\,,
	&\;\;
	\tilde{v}^{({\cal S})}(\varphi)
	\left|k-k_{1F}^{({\cal S})}\right|<\varepsilon_0,\\
	0\,,&\;\;\text{otherwise}\,,
\end{array}\right.
\end{equation}
where
$\varepsilon_0 \sim W_{\rm SDW}$
is the cutoff energy.
In the limit of
$T\to T_c$
one can linearize the equation for the superconducting order parameter
taking
$\Delta^{({\cal S}')}_{\mathbf{p}'}=0$
in the square roots in the integrals in
Eq.~\eqref{GapSCeq}.
Keeping in mind all aforementioned formulas and taking the integral over
$k$ in the limit
$\varepsilon_0/T\gg1$,
we obtain the equations for
$\Delta^{({\cal S})}(\varphi)$
in the form
\begin{equation}
\label{DeltaSCphi}
\Delta^{({\cal S})}(\varphi)=-\sum_{{\cal S}'}\int\limits_0^{2\pi}d\varphi'\frac{\tilde{\Gamma}^{({\cal S},{\cal S}')}_{\varphi\varphi'}k_{2F}^{({\cal S}')}\Delta^{({\cal S}')}(\varphi')}
{v_{\rm BZ}|\tilde{v}^{({\cal S}')}(\varphi')|}\ln\frac{E^*}{T}\,,
\end{equation}
where
$E^*=\varepsilon_0/(2A)$
and
$\ln A=\ln\pi/4-C$
(where $C$ is the Euler's constant,
$A\cong0.441$).

We calculate the functions
$\tilde{\Gamma}^{({\cal S},{\cal S}')}_{\varphi\varphi'}$
in
Eq.~\eqref{Gamm_SS}
numerically. An appropriate choice of the phase of the wave functions
$\Phi_{\mathbf{pG}is\sigma}^{({\cal S})}$
makes
$\tilde{\Gamma}^{({\cal S},{\cal S}')}_{\varphi\varphi'}$
real. The dependence of
$\tilde{\Gamma}^{({\cal S},{\cal S}')}_{0\varphi'}$
on
$\varphi'$
calculated for
$n=-1.75$
is shown in
Fig.~\ref{FigGamma}. We see that the absolute value of
$\tilde{\Gamma}^{({\cal S},{\cal S}')}_{0\varphi'}$
has maxima at
$\varphi'=\pi$
if
${\cal S}\neq{\cal S}'$,
see panel (b). The maxima of
$\tilde{\Gamma}^{({\cal S},{\cal S}')}_{0\varphi'}$
for
${\cal S}={\cal S}'$
are located near the
$\varphi'=\pi$,
panel (a). When
$\varphi\neq 0$,
the functions
$\tilde{\Gamma}^{({\cal S},{\cal S}')}_{\varphi\varphi'}$
have maxima at
$\varphi'\approx\varphi+\pi$.
Such a behavior of
$\tilde{\Gamma}^{({\cal S},{\cal S}')}_{\varphi\varphi'}$
can stabilize the superconducting state just due to the electron repulsion.
To show this, let us choose the trial function for
$\Delta^{({\cal S})}(\varphi)$
in the form
$\Delta^{({\cal S})}(\varphi)=\Delta^{({\cal S})}_0\cos\varphi$.
By multiplying both sides of
Eq.~\eqref{DeltaSCphi}
by
$2\cos\varphi$
and integrating over $\varphi$ one obtains the equation for
$\Delta^{({\cal S})}_0$:
\begin{equation}
\label{GapSCeqLambda}
\Delta^{({\cal S})}_0=\sum_{{\cal S}'}\lambda^{({\cal S},{\cal S}')}\Delta^{({\cal S}')}_0\ln\frac{E^*}{T}\,,
\end{equation}
where
\begin{equation}
\label{Lambda}
\lambda^{({\cal S},{\cal S}')}
=
-\frac{a^2\sqrt{3}k_{2F}^{({\cal S}')}}{2\pi}
\!\!\int\limits_0^{2\pi}\!\!\frac{d\varphi}{2\pi}\!\!
\int\limits_0^{2\pi}\!\!\frac{d\varphi'}{2\pi}
\frac{
	\tilde{\Gamma}^{({\cal S},{\cal S}')}_{\varphi\varphi'}
	\cos\varphi\cos\varphi'
}{
	|\tilde{v}^{({\cal S}')}(\varphi')|
}\,.
\end{equation}
The most important is that the double integral in
Eq.~\eqref{Lambda}
is negative for
${\cal S}={\cal S}'$
and
$\lambda^{({\cal S},{\cal S})}>0$
due to the properties of
$\tilde{\Gamma}^{({\cal S},{\cal S}')}_{\varphi\varphi'}$
described above. As a result,
Eq.~\eqref{GapSCeqLambda}
has non-trivial solutions for two values of $T$, and the maximum of these
two temperatures corresponds to
$T_c$.
The result can be presented in the form
\begin{equation}
\label{Tc}
T_c=E^*e^{-1/\Lambda},
\end{equation}
where
\begin{equation}
\Lambda\!=\!\frac{\left[\lambda^{(3,3)}+\lambda^{(4,4)}+\sqrt{[\lambda^{(3,3)}-\lambda^{(4,4)}]^2+4\lambda^{(3,4)}\lambda^{(4,3)}}\right]}{2}.
\end{equation}

We calculate $\Lambda$ numerically for three doping levels corresponding to
the Fermi surfaces shown in
Fig.~\ref{FigFS}.
For
$n=-1.75$,
$n=-1.69$,
and
$n=-1.67$
we obtain, respectively,
$\Lambda=0.05$,
$\Lambda=0.09$,
and
$\Lambda=0.08$.
Thus, the maximum
$T_c$
corresponds to
$n=-1.69$,
Fig.~\ref{FigFS}(b).
Taking for estimate
$E^*=W_{\text{SDW}}\cong17$\,meV,
we obtain in the latter case
$T_c\cong2.6$\,mK.
This value is much smaller than the experimentally
observed~\cite{NatureSC2018,MottSCNature2019}
$T_c\approx 1.7$\,K.
Thus, the considered Coulomb interaction alone is not enough to stabilize
the superconducting state with experimentally observed critical
temperature. The implications of this finding are discussed below.

\section{Discussion and Conclusions}
\label{Conclusions}

In this paper we consider a possibility of superconducting phase in MAtBLG,
and, more specifically, a Coulomb-interaction-driven superconducting
mechanism in MAtBLG. At the center of our proposal is the notion that at
least some parts of the MAtBLG Fermi surface remain ungapped despite the
SDW order parameter presence. The fermionic degrees of freedom that remain
at the Fermi energy even after the emergence of the SDW order is a peculiar
feature of
MAtBLG~\cite{OurtBLGPRB2019, Nematic, OurJETPLetters2022}.
The residual Fermi surface can host a weaker order parameter, such as a
superconductivity. This is the most important theoretical point of our
proposal.

This scenario has three obvious consequences, which can be tested
experimentally. (i)~The superconductivity coexists with the (stronger)
nematic SDW phase, (ii)~the superconducting order parameter is unavoidably
nematic, inheriting its nematicity from the underlying SDW order parameter,
and (iii)~our proposal entails large coherence length $\xi$: the usual BCS
estimate
$\xi = v_{\rm F}/\Delta \sim LW_{\rm SDW}/\Delta$
suggests that $\xi$ greatly exceeds the moir{\'{e}} period $L$, which
itself is significant, due to large ratio
$W_{\rm SDW}/\Delta$.
Note also that, due to (i) and (ii), a familiar classification of
superconducting order parameters into $s$-, $p$-, and $d$-wave symmetry
classes is impossible.

Nematic features of both the low-temperature superconducting phase and the
higher-temperature non-superconducting ``metallic'' state were indeed
experimentally
detected~\cite{cao2021nematicity}.
This finding is consistent with (i) and (ii) above.
Observed transport anisotropy of ``metallic" phase (see Fig.~3b of
Ref.~\onlinecite{cao2021nematicity})
is qualitatively consistent with a nematic Fermi
surface~\cite{OurtBLGPRB2019,Nematic,OurJETPLetters2022}
plotted in our
Fig.~\ref{FigFS}.
Namely, one can infer from this figure that the transport remains anisotropic
as long as the SDW order parameter is not destroyed by temperature.

While detailed description of the SDW phase is beyond the scope of this
manuscript, we will make the following two comments. In
Ref.~\onlinecite{cao2021nematicity}
the ratio
$\alpha = (R_1 - R_2)/(R_1 + R_2)$,
where
$R_{1,2}$
represent the resistivity tensor eigenvalues, serves as an experimental
measure of transport anisotropy. The absolute value of
$\alpha = \alpha (T,n)$
is between zero (for purely isotropic cases) and unity (for extreme
anisotropy). The data unambiguously indicate that the resistivity is
anisotropic both in the superconducting and deep in the ``metallic'' phases, but
$|\alpha |$
remains quite small for most $T$ and $n$ values, begging the question of
how this smallness fits into the discussed theoretical framework.
Considering this issue, one must keep in mind that, as the Supplemental
Material to
Ref.~\onlinecite{cao2021nematicity}
explains, the performed measurement always underestimates $|\alpha|$. Additionally,
in our model the Fermi surface anisotropy is quite moderate, as
Fig.~\ref{FigFS}
indeed attests, implying moderately low $|\alpha |$.
Consequently, we interpret the results of
Ref.~\onlinecite{cao2021nematicity}
as being qualitatively consistent with our scenario.

The second comment is related to
Ref.~\onlinecite{nature2023},
which presents an STM study of MAtBLG. This investigation, unlike previous
papers~\cite{MottNematicNature2019,KerelskyNematicNature2019},
did not report a nematic phase of MAtBLG. Although at this stage a
confident resolution of this discrepancy is impossible, we can hypothesize
that it may be a manifestation of the competition between multiple
dissimilar low-energy phases in the considered
system~\cite{Nematic}.
If the energies of the competing phases are sufficiently close, the outcome
of the competition is determined by an interplay of a number of poorly
controlled factors unique to a specific MAtBLG device. In this framework,
it becomes natural that several seemingly identical samples demonstrate
different low-temperature properties.

Besides the presence of the Fermi surface, an essential ingredient of a
mechanism is a source of attraction keeping Cooper pairs together. In the
previous section we attempted to assess to which extent the renormalized
Coulomb interaction can serve this purpose. Our calculations revealed that
the resultant critical temperature is much lower than the value observed in
the experiment.

Clearly, the discrepancy in terms of
$T_c$
requires additional analysis. It is easy to convince oneself that the root
cause of the superconducting instability weakness is the weakness of the
coupling constant $\Lambda$. In our estimates $\Lambda$ never exceeded 0.1,
making the BCS exponent
$\exp( - 1/ \Lambda)$
extremely small.

Moreover, in the regime of small $\Lambda$, any inaccuracy in $\Lambda$ is
greatly amplified by the BCS exponential function. To illustrate this
sensitivity in our circumstances, let us increase the coupling constant
two-fold, from 0.09 to 0.18. Then the critical temperature grows by more
than two orders of magnitude, from
2.6\,mK
to
0.66\,K,
which compares favorably against
1.7\,K
measured experimentally. This simple calculation reminds us that an
order-of-magnitude estimate of $\Lambda$ is insufficient for an
order-of-magnitude estimate of
$T_c$.
This issue is particularly pressing in the limit of low $\Lambda$, as in
our case.

We envision two possibilities that can reconcile the theory with the
experiment. One option is simply to resign to the fact that approximate
nature of our calculations limits us to order-of-magnitude estimate
$\Lambda$, which is equivalent to order-of-magnitude estimate of
$\ln T_c$.
We should not consider this viewpoint as excessively defeatist. After all,
any many-body calculation is performed under numerous assumptions that
skew the final answer. For MAtBLG the situation is worsened by lack of
reliable knowledge about the interlayer tunneling.

Alternatively, we can add phonons to our mechanism. One can imagine two
possibilities for phonon-mediated attraction. (i)~The phonons increase the
coupling constant $\Lambda$ discussed in the previous section, increasing
the critical temperature. (ii)~On the other hand, the phonon-mediated
attraction may stabilize a superconducting order parameter of different
type (e.g., nodeless). In the latter case, the competition between two (or
more) order parameters of different structures becomes a possibility.

The superconductivity in MAtBLG is experimentally observed
both below and above half-filling. In our paper we present the results for
doping level slightly above $n=-2$. Similar calculations can be done for
$n<-2$.
In that case (when doping $n$ is not very far from half-filling)
the Fermi surface structure consists of two closed curves centered at
$\bm{\Gamma}$
point. These Fermi surface sheets are not circular but elongated along the
vector
$\mathbf{G}_2$.
When evaluating the superconducting coupling constant $\Lambda$ for some
doping levels with
$n<-2$
one discovers that $\Lambda$ is of the same order as in the case of
$n>-2$.
For example, for
$n=-2.5$
we obtain
$\Lambda\approx0.1$.
Thus, our theory predicts the same superconducting temperatures for two
superconducting domes near the half-filling.

Finally, when interpreting experimental data for MAtBLG, it is necessary to
remember that the system might experience electronic phase separation. For
twisted bilayer graphene this phenomenon has been discussed in
Ref.~\onlinecite{OurPhaSepJETPL2020},
but it itself is not uncommon in theoretical models for doped SDW
phase~\cite{tokatly1992,igoshev2010,aa_graph2013,phasep_pnics2013,
graph_phasep2013,nesting_review2017,ourPRB_phasepAFM2017,
pressure_phasep_rakhmanov2020,kokanova2021hubbard_inhomogen},
as well as for other continuous phase transitions affected by
doping~\cite{Fine_Egami2008phase_sep}.
Phase separation frustrated by long-range Coulomb interaction may lead to
spatial pattern formation altering
transport~\cite{Narayanan2014q1d_exper}
and other physical properties of a sample.

In conclusion, we argued that MAtBLG can enter a superconducting phase
coexisting with the SDW-like ordering. The mean field description of the
host SDW state accounts for on-site, and both in-plane and out-of-plane
nearest-neighbor intersite anomalous expectation values. Numerical mean
field minimization reveals that the SDW order leaves small multi-component
Fermi surface ungapped. Near the half-filling the SDW order parameters
partially break the MAtBLG point symmetry group that leads to the Fermi
surface nematicity. For superconductivity the presence of the ungapped
Fermi surface is crucial as it bypasses the competition between the
magnetic and superconducting phases, which the (much weaker)
superconductivity cannot win. Additionally, we explore the possibility of
purely Coulomb-based mechanism of the superconductivity in MAtBLG. The
screened Coulomb interaction is calculated within the random phase
approximation. We show that near the half-filling the renormalized Coulomb
repulsion indeed stabilizes the superconducting state. The superconducting
order parameter has two nodes on the Fermi surface. We estimate the
superconducting transition temperature and discuss the implications of our
proposal.

\begin{acknowledgments}
This work is supported by RSF grant No.~22-22-00464,
\url{https://rscf.ru/en/project/22-22-00464/}. We acknowledge the Joint
Supercomputer Center of the Russian Academy of Sciences (JSCC RAS) for the
computational resources provided.
\end{acknowledgments}



\end{document}